\documentclass[11pt]{article}	
\pdfoutput=1
\usepackage{jcappub}

\usepackage{natbib}

\usepackage{amsmath,amsfonts,amssymb}
\usepackage{setspace}
\usepackage{tocloft}

\usepackage[normalem]{ulem}
\usepackage{color}

\def\wisk#1{\ifmmode{#1}\else{$#1$}\fi}
\def\deg    {\wisk{^\circ}}

\def\asec   {\wisk{^{\prime\prime}\ }}

\def\lsim   {\wisk{_<\atop^{\sim}}}

\def\aap {Astronomy and Astrophysics}
\def\apj {The Astrophysical Journal}
\def\apjl {The Astrophysical Journal Letters}
\def\apjs {The Astrophysical Journal Supplement Series}
\def\jatis {Journal of Astronomical Telescopes, Instruments, and Systems}
\def\jcap {Journal of Cosmology and Astroparticle Physics}
\def\prd {Physical Review D}

\title{Systematic error mitigation for the PIXIE Fourier transform spectrometer \\
}

\author[a]{A. Kogut}
\author[b]{Dale Fixsen}
\author[c]{Nabila Aghanim}
\author[d]{Jens Chluba}
\author[e]{David T. Chuss}
\author[f]{Jacques Delabrouille}
\author[g]{Brandon S. Hensley}
\author[h,i]{J. Colin Hill}
\author[j]{Bruno Maffei}
\author[k,i]{Anthony R. Pullen}
\author[d]{Aditya Rotti}
\author[a]{Eric R. Switzer}
\author[a]{Edward J. Wollack}
\author[l]{Ioana Zelko}

\note{Corresponding author.}

\affiliation[a]{NASA Goddard Space Flight Center, 8800 Greenbelt Road, Greenbelt, MD, 20771, USA}
\affiliation[b]{Department of Astronomy, CRESST II, University of Maryland, College Park MD 20740 USA}
\affiliation[c]{Universit\'e Paris-Saclay, CNRS, Institut d'Astrophysique Spatiale, B\^atiment 121, 91405 Orsay, France}
\affiliation[d]{Jodrell Bank Centre for Astrophysics, Department of Physics and Astronomy, The University of Manchester, Manchester M13 9PL, UK}
\affiliation[e]{Department of Physics, Villanova University, 800 Lancaster Avenue, Villanova, PA 19085, USA}
\affiliation[f]{CNRS-UCB International Research Laboratory, Centre Pierre Bin\'etruy,  IRL 2007, CPB-IN2P3, Berkeley, CA 94720, USA }
\affiliation[g]{Department of Astrophysical Sciences, Princeton University, Princeton, NJ 08544, USA}
\affiliation[h]{Department of Physics, Columbia University, New York, NY 10027, USA}
\affiliation[i]{Center for Computational Astrophysics, Flatiron Institute, 162 Fifth Avenue, New York, NY 10010, USA}
\affiliation[j]{Institut d'Astrophysique Spatiale, CNRS-Universit\'e Paris-Saclay, Orsay, 91405, France}
\affiliation[k]{Center for Cosmology and Particle Physics, Department of Physics, New York University, 726 Broadway, New York, NY, 10003, USA}
\affiliation[l]{Canadian Institute for Theoretical Astrophysics, University of Toronto, 60 St George Street, Toronto, M5S 3H8, Ontario, Canada}

\emailAdd{Alan.J.Kogut@nasa.gov}

\abstract{
The Primordial Inflation Explorer (PIXIE)
is an Explorer-class mission concept
to measure
the spectrum and polarization of the cosmic microwave background.
Cosmological signals are small compared to the
instantaneous instrument noise,
requiring strict control of instrumental signals.
The instrument design provides
multiple levels of null operation,
signal modulation,
and signal differences,
with only few-percent systematic error suppression
required at each level.
Jackknife tests
based on discrete instrument symmetries
provide an independent means to
identify, model, and remove remaining instrumental signals.
We use detailed time-ordered simulations,
including realistic performance and tolerance parameters,
to evaluate the instrument response
to broad classes of systematic errors
for both spectral distortions and polarization.
The largest systematic errors
contribute additional white noise
at the few-percent level 
compared to the dominant photon noise.
Coherent instrumental effects which do not integrate down
are smaller still,
and remain several orders of magnitude
below the targeted cosmological signals.
}


\notoc

\begin{document}
\maketitle
\flushbottom


\begin{spacing}{1}   		

\section{Introduction}
Measurements of the cosmic microwave background (CMB) 
have played a central role in the development of modern cosmology.
The blackbody form of the CMB spectrum 
provides the foundation for the hot big bang model,
tracing the currently-observed expansion of the universe
back in time 
to a hot, dense phase in close thermal equilibrium.
Deviations from a perfect blackbody
follow the subsequent thermal history
as the universe evolves to its present form,
providing new tests of the standard cosmological model
\cite{kogut/etal:2019, 
chluba/etal:2021}.
Spatial maps of small temperature perturbations
about the blackbody mean
track density perturbations,
providing detailed information
on the geometry, constituents, and evolution of the universe
\cite{wmap_parameters_2013,
planck_parameters_2018}.
Maps of CMB polarization
yield insight into the epoch of reionization,
the growth of structure in the low-redshift universe,
and
offer the tantalizing possibility 
to detect the signature of an inflationary epoch in the early universe
\cite{cmb_s4_2019,
LiteBIRD_2019,
pico_2019}.

\begin{figure}[b]
\centerline{
\includegraphics[width=5.0in]{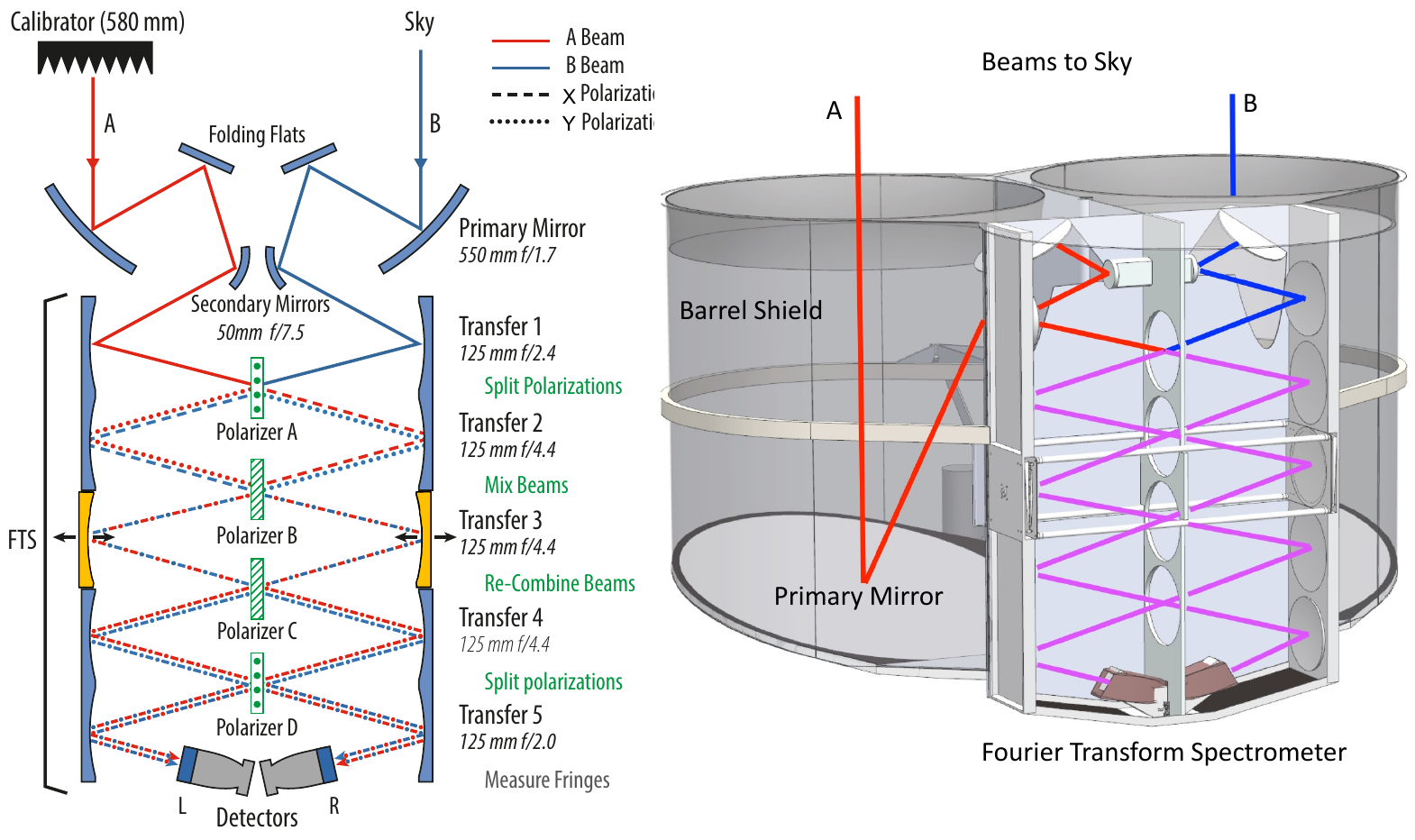}}
\caption{
Schematic rendering of the PIXIE instrument,
showing the optical path (left)
and the beam-forming optics (right).
The moving phase-delay mirror pair is highlighted.
The Fourier transform spectrometer
interferes two co-pointed beams
while maintaining all elements within 5~mK
of the CMB monopole temperature.
The resulting null operation,
differential measurement,
and complex signal modulation
mitigate broad classes of systematic error.
}
\label{pixie_optics_fig}
\end{figure}

Several decades of instrument development
have produced orders-of-magnitude improvement
in sensitivity,
from the $\mu$K sensitivity
of the Cosmic Background Explorer
\cite{bennett/etal:1996}
to the nK sensitivity
proposed for missions in the near future
\cite{cmb_s4_2019,
LiteBIRD_2019,
pico_2019}.
As raw sensitivity improves,
careful control of systematic errors are required,
as well as means to detect, identify, and correct 
residual instrumental signatures.

The Primordial Inflation Explorer (PIXIE)
is a proposed space mission
to measure the frequency spectrum and polarization
of the CMB
and astrophysical foregrounds
on angular scales of 1\deg ~and larger
\cite{kogut/etal:2011,
kogut/fixsen:2019,
pixie_mission:2023}.
Unlike most CMB instruments,
which focus incident light onto
an array of diffraction-limited detectors
each operating within a photometric passband 
$\Delta \nu / \nu \approx 0.2$,
PIXIE interferes two co-pointed beams
within a polarizing Fourier Transform Spectrometer (FTS)
to measure the autocorrelation function 
of the sky signal.
Although targeting comparable sensitivity
as other CMB missions,
PIXIE's different instrumental configuration
produces a correspondingly different response 
to potential systematic errors
in both the frequency and spatial domains.
Previous papers have described
the PIXIE data pipeline
and
systematic error mitigation
for certain classes of instrumental effects
\cite{nagler/etal:2015,
pixie_4_port_2019,
naess/etal:2019}.
In this paper,
we extend this treatment
to quantify the systematic error mitigation
in both the frequency and spatial domains
and discuss
the power of jackknife tests
to identify and remove remaining instrumental signals.

\section{PIXIE Instrument and Observations}

Figure \ref{pixie_optics_fig} shows the PIXIE instrument concept.
It consists of a polarizing Fourier transform spectrometer
with two input ports illuminated by co-pointed beams on the sky.
The 55~cm diameter primary mirrors produce circular tophat beams
with diameter 2.65\deg,
roughly equivalent to Gaussian beams
with 1.65\deg~full width at half maximum.
A pair of polarizing wire grids
splits the light from each beam
into orthogonal linear polarizations
then mixes the beams.
A movable structure with two mirrors
introduces an optical phase delay
before a second set of polarizing grids
recombines the beams
and routes them to the two output ports.
Each output port contains 
a pair of polarization-sensitive bolometers,
mounted to absorb mutually orthogonal polarization states.

\begin{figure}[b]
\centerline{
\includegraphics[width=5.0in]{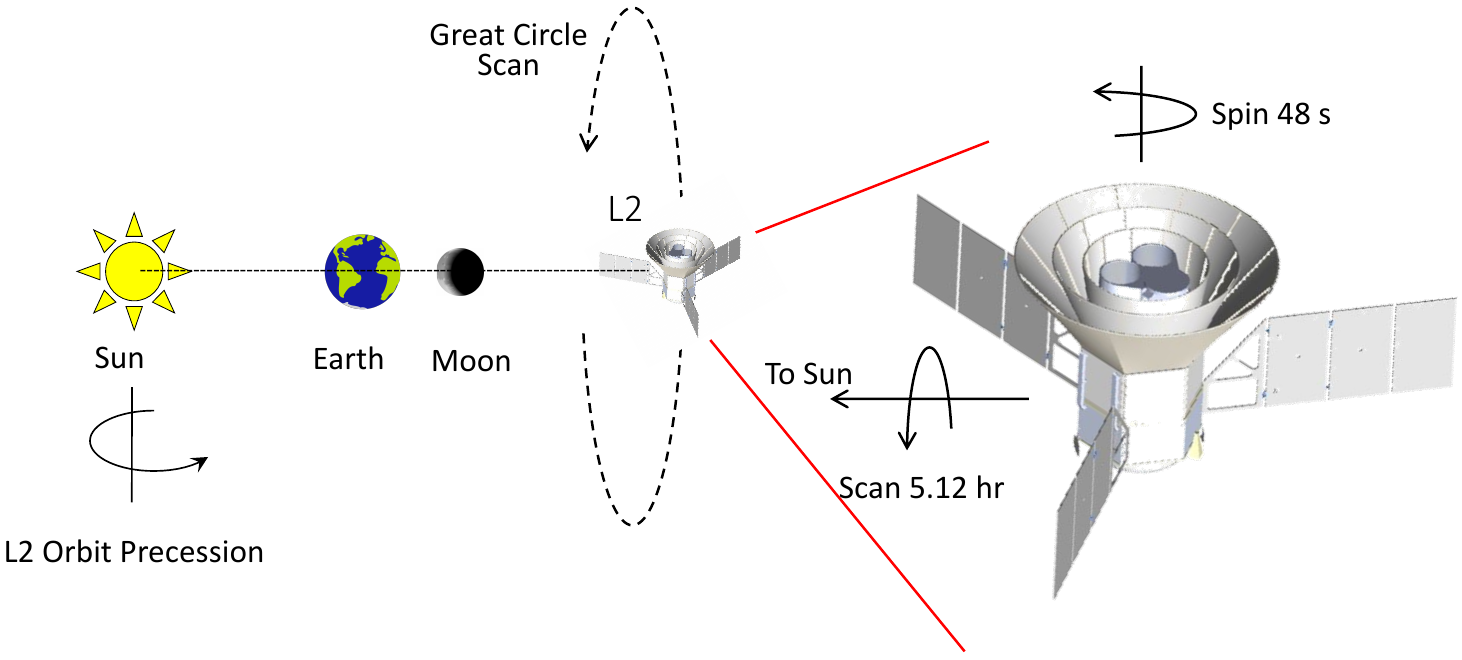}}
\caption{
PIXIE will observe at the Sun-Earth L2 point.
The spacecraft spins about the beam boresight
while simultaneously scanning the beams
in a great circle perpendicular to the sun line.
The scan pattern maps the full sky every 6 months.
}
\label{pixie_scan}
\end{figure}

As the phase-delay mirrors sweep back and forth,
each of the 4 detectors samples the resulting 
interference fringe pattern 
as a function of the optical phase delay.
Let $\vec{E} = E_x \hat{x} + E_y \hat{y}$ 
represent the electric field incident from the sky
in an orthogonal basis 
$\hat{x}$ and $\hat{y}$.
The power $P$ at the detectors
as a function of the phase delay $z$
may be written
\begin{eqnarray}
P_{Lx} &=& 1/2 ~\smallint \epsilon_{Lx} \{ ~(f_A E_{Ax}^2+f_B E_{By}^2)+(f_A E_{Ax}^2-f_B E_{By}^2) \cos(z\omega /c) ~\}d\omega   \nonumber \\
P_{Ly} &=& 1/2 ~\smallint \epsilon_{Ly} \{ ~(f_A E_{Ay}^2+f_B E_{Bx}^2)+(f_A E_{Ay}^2-f_B E_{Bx}^2) \cos(z\omega /c) ~\}d\omega   \nonumber \\
P_{Rx} &=& 1/2 ~\smallint \epsilon_{Rx} \{ ~(f_A E_{Ay}^2+f_B E_{Bx}^2)+(f_B E_{Bx}^2-f_A E_{Ay}^2) \cos(z\omega /c) ~\}d\omega    \nonumber \\
P_{Ry} &=& 1/2 ~\smallint \epsilon_{Ry} \{ ~(f_A E_{Ax}^2+f_B E_{By}^2)+(f_B E_{By}^2-f_A E_{Ax}^2) \cos(z\omega /c) ~\}d\omega~,
\label{full_p_eq}
\end{eqnarray}
where
$x$ and $y$ refer to orthogonal linear polarizations,
$\epsilon$ is the (polarized) detector absorption efficiency,
$f$ is the transmission through the optics to the detector,
L and R refer to the detectors in the left and right concentrators,
A and B refer to the two input beams,
and $\omega$ is the angular frequency of incident radiation.
The optical phase delay $z$ is related to the physical mirror position $\Delta L$ as
\begin{equation}
z = 4 \cos(\theta) \cos(\delta/2) \Delta L ~,
\label{mirror_eq}
\end{equation}
where
$\theta$ is the angle of incident radiation
with respect to the mirror movement,
$\delta$ is the dispersion in the beam,
and the factor of 4 
reflects the symmetric folding of the optical path.
The factor of $1/2$ rather than $1/4$ for each of the four detectors
results from use of 2 input ports
rather than a single port.
When both input ports are open to the sky,
the power at each detector consists of a dc term 
proportional to the intensity 
$E_x^2 + E_y^2$
(Stokes $I$)
plus a term 
modulated by the phase delay $z$,
proportional to the linear polarization 
$E_x^2 - E_y^2$
(Stokes $Q$) 
in instrument-fixed coordinates.
Rotation of the instrument about the beam axis
rotates the instrument coordinate system
relative to the sky
to allow separation of Stokes $Q$ and $U$ parameters on the sky.
Since the instantaneous measurement of the Stokes $Q$ parameter
is defined with respect to the instrument,
the mapping of $Q$ and $U$ to sky coordinates
can be performed in any fixed sky coordinate system
(e.g. celestial, ecliptic, or Galactic coordinates).
A full-aperture blackbody calibrator can be deployed
to block either of the two input ports,
or stowed so that both ports view the sky.
With one input port terminated by the calibrator,
the modulated term is then proportional to the
difference between the sky signal 
and the calibrator,
providing sensitivity to the sky signal in Stokes $I$, $Q$, and $U$
as well as a known reference signal for calibration.
Circular polarization (Stokes $V$)
from astrophysical sources
is negligible at millimeter and sub-mm wavelengths,
and only appears in the imaginary part of the PIXIE Fourier transform
($\S$3.2.5).

Figure \ref{pixie_scan} shows the observing strategy.
PIXIE will be located at the second Sun-Earth Lagrange point (L2).
The spacecraft spins about the instrument boresight
while slowly scanning the boresight
in a great circle perpendicular to the sun line
so that the full sky is mapped evey 6 months.
For ease of analysis,
the mirror stroke,
spacecraft spin,
and great-circle scan
are maintained in a fixed ratio
at widely-separated time scales,
driven by a single master clock to avoid beat frequencies.
The nominal periods of
4 second mirror stroke,
48 second spin,
and 18432 second (5.12 hour) scan
provide
12 mirror strokes per spin
and
384 spins per great-circle scan.
Maintaining fixed ratios
allows clean separation 
between the Fourier transforms
producing the frequency spectra (mirror stroke),
Stokes parameters (spin),
and spatial mapping (scan)
but is not required for mission success.

The mission data
consists of the power on each of the 4 detectors
sampled at 256~Hz,
supplemented by
ancillary data
including mirror position,
instrument temperatures,
and spacecraft pointing.
The sampled fringe patterns
sorted by mirror position
(interferograms, or IFGs)
form the basic time-ordered data.
The interferograms for each detector
are sorted by
spin angle and sky pixel,
then
Fourier transformed with respect to
mirror position, spin angle, and sky coordinates
to derive 
frequency spectra 
for each of the IQU Stokes parameters
in each spatial pixel.
With HEALPIX pixelization at resolution \texttt{NSIDE=64}
\cite{healpix/2005},
the resulting sky cube has
49152 spatial pixels of diameter 0.9\deg,
each with data from at least 200 independent mirror strokes.
Over a 2-year mission,
the integration time per pixel
ranges from 800 seconds on the ecliptic equator
to 50,000 seconds at the north and south ecliptic poles,
with a typical mid-latitude pixel
observed for 940 seconds.

The frequency resolution 
within each spatial pixel
depends on the mirror stroke and apodization.
The nominal mirror stroke length varies
to produce either
512 frequency bins of width 19.2 GHz
for spectral distortion measurements (calibrator blocking one beam)
or
256 bins of width 38.3~GHz
for polarization measurements (calibrator stowed)
\cite{pixie_mission:2023}.
Scattering filters on the beam-forming optics
limit the optical response
so that channels at frequencies above $\sim$6~THz
contain no sky signal.

Comparison of the sky to the blackbody calibrator
provide the necessary calibration
from digitized telemetry units
to surface brightness on the sky
\cite{kogut/fixsen:2019}.
The instrument noise is dominated by photon statistics
from CMB monopole and blackbody calibrator,
each at 2.725~K.
Photon noise 
at sub-mm wavelengths
from higher-temperature sources
(interstellar dust, zodiacal light, and the cosmic infrared background)
contributes less than 20\% to the total noise
for pixels at Galactic latitude $|b| > 20\deg$.

\begin{figure}[t]
\begin{center}
\includegraphics[width=3.0in]{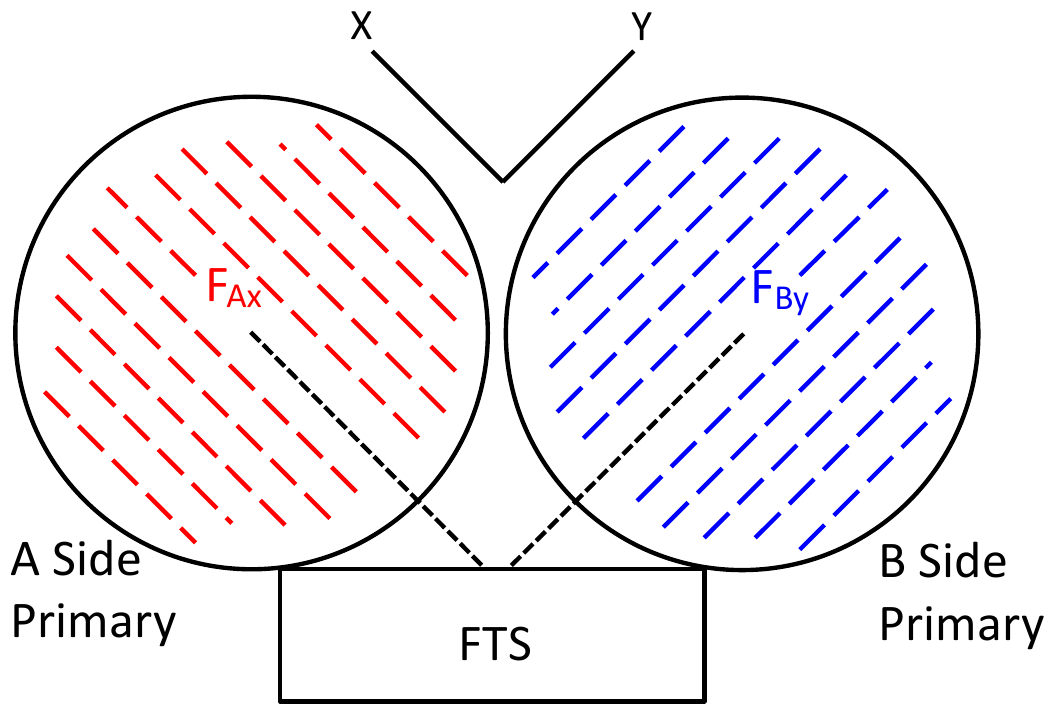}
\end{center}
\caption{ 
Schematic of the PIXIE optical system 
viewed from the beam axis,
showing the principal instrument symmetry.
By construction, the
$\hat{x}$ polarization on the A side
is simply the mirror reflection of the
$\hat{y}$ polarization on the B side.
\label{symmetry_cartoon}
}
\end{figure}

Instrument symmetry plays a major role
in suppressing systematic error.
The A and B sides of the instrument
are symmetric about the midline
between the two primary mirrors
(Fig. \ref{symmetry_cartoon}).
By construction, the
$\hat{x}$ polarization on the A side
is simply the mirror reflection of the
$\hat{y}$ polarization on the B side.
Machining and assembly tolerances
break this symmetry
and can source systematic error.
Section \ref{syserr_section}
shows how the instrument symmetry and Fourier transform
minimize potential systematic error,
while
Section \ref{jackknife_section}
discusses jackknife tests
to identify and correct residual errors.

\section{Systematic Error}
\label{syserr_section}

Systematic errors may be defined as effects
from the instrument and data processing
which create organized differences
between the true sky 
and its representation in data.
An extensive literature analyzes common sources of systematic error
and mitigation strategies for CMB measurements
\cite{
dmr_syserr_1992,
firas_syserr_1994,
kaplan:2001,
hu/etal:2003,
bunn:2006,
shimon/etal:2008,
singal/etal:2011,
nagler/etal:2015,
planck_lfi_syserr_2016,
planck_hfi_syserr_2016,
pixie_4_port_2019,
naess/etal:2019}.
A useful taxonomy distinguishes between
additive effects
(which exist independent of specific sky signals)
and
multiplicative effects
(which modulate sky signals
but vanish if the underlying sky signal is not present).
Examples of the former include
post-detection electronics pickup,
instrumental emission, and $1/f$ noise, 
while the latter includes effects such as
gain error, bandpass error, and beam effects.

Several strategies have proven effective 
for the identification, suppression, and mitigation of systematic error.
Null operation maintains the instrument near the same temperature as the sky,
minimizing offsets generated by internal absorption, emission, or reflection.
Differential operation compares signals to cancel common-mode emission
from unwanted sources
while 
reducing the effect of post-detection gain fluctuations.
Signal modulation imposes a pre-determined time-ordered variation 
on  specific input signals;
subsequent matched demodulation
separates the desired signals from other (unmodulated) sources.
Jackknife tests --
the comparison of equivalent data sets
sorted by some external factor --
provide a blind method to search for residuals
down to the level of the noise in the selected data.
While PIXIE employs all of these methods,
we focus below on
the role of the FTS in systematic error suppression.

\subsection{Additive Effects}

Additive effects are independent of the sky signal.
We review common additive effects
and use simulations
to assess their impact on the PIXIE data.

\subsubsection{Additive Periodic Signals
\label{periodic_section}}

Signals that repeat with a fixed period
commensurate with the sky scan
can build up to observable levels.
Common examples are 
electronic pickup of signals 
at harmonics of the beam motion across the sky
or (for polarization)
harmonics of the spin period or other polarization modulation.
We use noiseless simulations to assess the extent
to which the FTS can suppress such signals.
We generate 6 months of time-ordered data
for a single detector,
consisting of a sine wave of unit amplitude
and fixed period,
and tag each time-ordered sample with the 
FTS phase delay,
spin angle,
and sky pixel.
We then project the time-ordered data into sky pixels
using the HEALPIX pixelization at resolution \texttt{NSIDE=64}.
With no FTS,
the simplest mapping scheme
bins the data from a single detector by spin angle and sky pixel
to project the sine signal
into full-sky maps in Stokes IQU parameters.
Including the FTS,
we sort the time-ordered data for a single detector
into individual interferograms,
perform the Fourier transform to the frequency domain,
and sort the synthesized spectra
by spin angle and sky pixel
to derive the IQU parameters in each voxel.

\begin{figure}[t]
\centerline{
\includegraphics[width=6.5in]{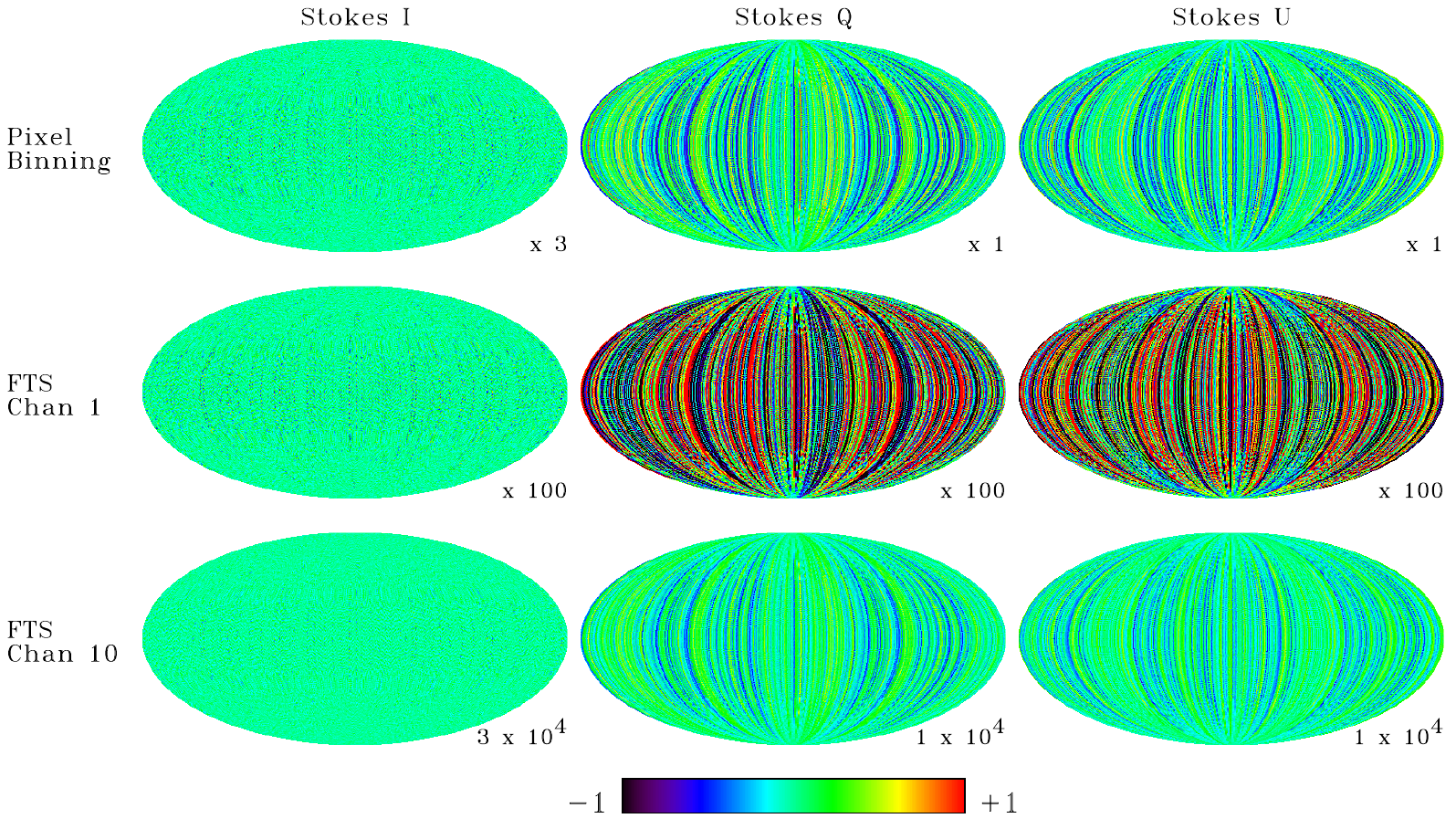}}
\caption{
Sky maps generated from an additive sinusoidal signal
of unit amplitude
at twice the spin frequency
are shown for the PIXIE scan pattern
(Mollweide projection in ecliptic coordinates).
Compared to simple pixel binning,
the FTS suppresses the projection into Stokes Q and U
with a steep frequency dependence.
Maps have been scaled in amplitude as noted
to allow a common color stretch from -1 to +1.
}
\label{spin_sine_maps}
\end{figure}

Figure \ref{spin_sine_maps}
shows the resulting IQU maps
for sinusoidal modulation
at twice the spin frequency.
As expected,
sorting the signal by sky pixel and spin angle
simply projects the spin-modulated signal 
into the Q and U polarization maps
at unit amplitude.
After Fourier transformation,
the Q and U maps show a similar spatial striping pattern,
but with amplitude reduced by a factor of 30 
in the lowest synthesized frequency channel
and 
falling as $\nu^{-5/2}$ in higher frequency channels.

\subsubsection{Internal Emission and Reflection}
\label{internal_emission}

Emission and reflection within the instrument 
can induce additive systematic error,
both from elements  
directly in the optical path
from the detectors to the sky 
(mirrors, grids)
as well as any stray-light view from the detectors 
to the walls of the instrument.
To minimize these effects,
the entire optical path --
primary mirrors, secondary mirrors, folding flat, FTS,
and all surrounding walls --
is maintained at temperatures within a few mK of the 2.725~K CMB monopole.
The largest possible instrumental signal is thus determined by the
few~mK maximum temperature difference between the instrument and the sky.
Isothermal operation alone reduces systematic errors
from internal absorption or reflection within PIXIE
(temperature differences 2.730 - 2.725 K)
by a factor of 250 compared to instruments with 4~K optics
(4 - 2.725 K),
and a factor of 50,000 compared to room-temperature optics
(295-2.725 K).

The double-differential design provides additional mitigation.
By Kirchoff's law,
sky photons absorbed by the mirrors or grids
are replaced by emitted photons
whose spectrum is well approximated
as a modified blackbody at the instrument temperature,
\begin{equation}
I_\nu = \kappa \left( \frac{\nu}{\nu_0} \right)^\alpha ~B_\nu(T_{\rm inst}) ~.
\label{inst_emission_1}
\end{equation}
For clean aluminum surfaces,
$\alpha = 1/2$
with
$\kappa = 0.002$ at reference frequency 
$\nu_0 = 100$~GHz
\cite{bock/etal:1995},
so that the error signal from any single surface
is of order 
$10^{-5} ~\partial B_\nu / \partial T$
where
$\partial B_\nu / \partial T$
is the temperature derivative of a blackbody spectrum at $T=2.725$~K.
The FTS then differences
the signal from each optical surface on the A side of the instrument
against the comparable signal
from the corresponding surface
on the B side.
After Fourier transformation,
the resulting systematic error signal
is proportional to the double difference of
the component emissivities and temperatures,
\begin{equation}
I_\nu = \Delta \kappa ~ 
\times \frac{\partial B_\nu}{\partial T} ~
\times \Delta T_{\rm inst} ~.
\label{inst_emission_2}
\end{equation}
With
$\Delta \kappa < 0.01 ~\kappa$
and
$\Delta T_{\rm inst} < 5$~mK,
the maximum error in the time-ordered data
is now of order 
$10^{-7} ~\partial B_\nu / \partial T$
\cite{nagler/etal:2015}.

Mission operations further mitigate the impact of differential emission.
If the temperature difference between optical components
on the A and B sides were constant in time,
the error signal (Eq. \ref{inst_emission_2})
would have peak amplitude of 100~Jy~sr$^{-1}$,
comparable to the cosmological signals.
Throughout the mission,
the temperature of each optical component
is individually controlled
to 100 $\mu$K accuracy
and 1 $\mu$K knowledge.
Twice each great-circle scan
(as the beams cross the north and south ecliptic poles)
all component temperatures are set
to new values
such that the time-ordered temperature profile
of any component
is orthogonal with any other component.
It is easily arranged
to spend half the mission with 
positive temperature difference 
$+ \Delta T_{\rm inst}$
between corresponding optical surfaces
and half with
negative difference
$- \Delta T_{\rm inst}$.
The systematic error thus cancels in the signal mean,
contributing only to the variance\footnote{
By appropriate choice of weights in the time domain,
the cancellation can be made nearly exact.}.
More generally,
we extend the sky map software
\cite{naess/etal:2019}
to include nuisance terms 
\begin{equation}
\kappa_i ~ T_{\rm inst, i}(t)
\label{nuisance_eq}
\end{equation}
proportional
to the temperature of each optical component,
effectively measuring 
the coupling to each optical surface.
After marginalization over the nuisance parameters,
the residual depends on the integrated noise 
$\delta T_{\rm therm}$
in the thermometer readout for each component,
\begin{equation}
\delta I_\nu = \Delta \kappa ~ 
\times \frac{\partial B_\nu}{\partial T} ~
\times \delta T_{\rm therm} ~.
\label{syserr_amp_stray_1}
\end{equation}
The thermometers are read out once per second
with read noise 1~$\mu$K.
Over the course of a 2-year mission,
the read noise for each component integrates down
to precision 
$\delta T_{\rm therm} \approx 0.1$~nK,
resulting in  residual error
$\delta I_\nu < 10^{-5}~$~Jy~sr$^{-1}$.
Since $T_{\rm inst}$ varies symmetrically about the sky temperature,
the nuisance signal is nearly orthogonal to the sky signal.
The addition of $\sim$20 global nuisance parameters
thus increases the noise 
in the fitted sky cubes
by a negligible amount.
Note that the correction only requires knowledge
of the temperature change in each component;
absolute thermometry is not required.

\begin{figure}[t]
\centerline{
\includegraphics[height=3.5in]{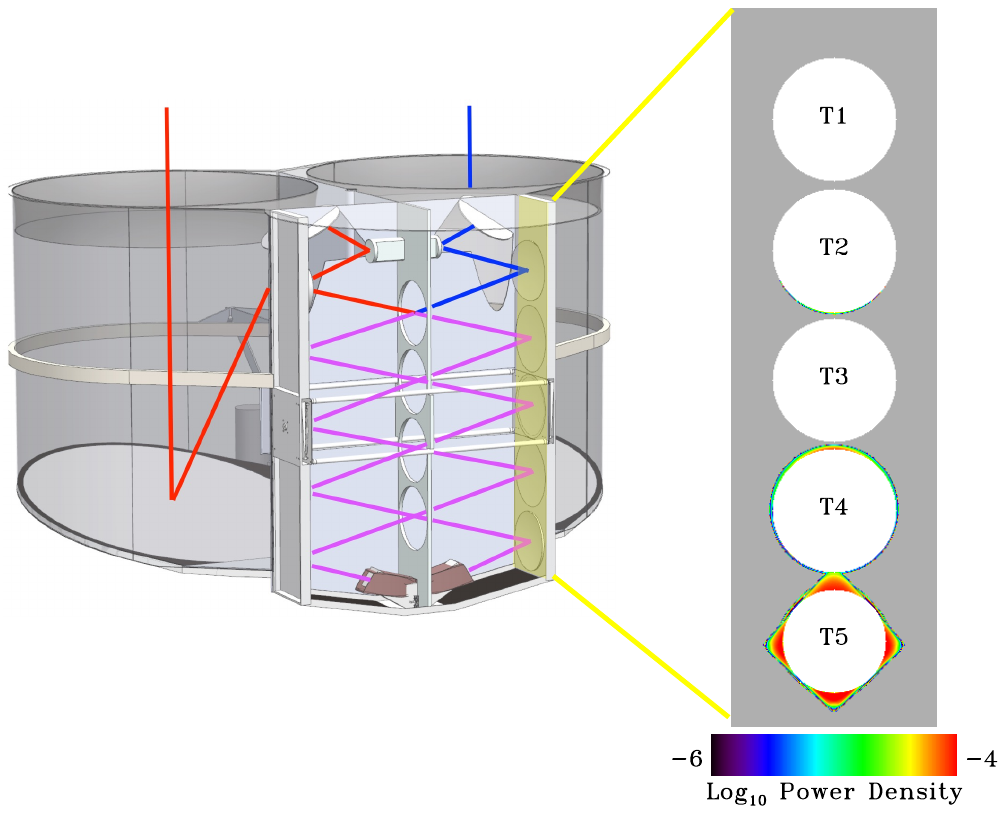}}
\caption{
Stray-light illumination pattern on the instrument wall
surrounding the FTS transfer mirrors.
Emission from components below the MTM 
(e.g. spillover at mirrors T4 and T5)
is not modulated by the FTS phase delay.
Temperatures of the mirrors, polarizers, and walls
are under active thermal control
and periodically varied
to separate stray light from sky signals.
}
\label{totem_wall}
\end{figure}

Emission from the instrument walls
can scatter into the beams
to reach the detectors.
We model the resulting stray-light signal
using a ray-trace code in the geometric optics limit.
For each linear polarization,
we generate a set of $10^9$ outgoing (skyward) rays,
each originating from a random position on the detector
and with a random skyward direction.
A ray-tracing algorithm
follows each ray
through the optical path.
Rays that miss the active optical elements
terminate on the walls of the instrument,
which are coated with a microwave absorber.
Figure \ref{totem_wall} shows the density of stray-light rays
terminating on the wall containing the transfer mirrors
within the FTS.
A circular cold stop at the 0.1~K horn mouth acts as a pinhole camera
to re-image the square base of the horn
onto the FTS exit port (mirror T5).
12\% of the rays exiting the concentrator
terminate on the wall surrounding mirror T5,
with an additional 2\% terminating at mirror T4.
Since these rays do not undergo the FTS phase delay,
they do not generate a fringe pattern
and instead  only appear 
as part of the unmodulated term in Eq. \ref{full_p_eq}.
As with the mirrors,
each wall is under active thermal control
and
maintained within a few mK of the sky temperature.
Un-modulated stray light may thus be treated
as an additive systematic error signal 
with maximum amplitude of 0.01\% of the total power
(dominated by the CMB monopole).
Since the time constants for individual temperature-controlled components
($\sim$10~seconds)
are slow compared to the 4-second mirror sweep,
component temperatures cannot vary rapidly
compared to individual interferograms.
The resulting stray-light signal 
from the wall surrounding Transfer Mirror 5
produces a systematic error $< 10^{-3}$~Jy~sr$^{-1}$.

Stray light originating from components 
skyward of Polarizer B
undergo beam-splitting and phase delay
and produce fringes at the detector.
The ray-trace analysis
shows that 
0.3\% of the beams terminate on these surfaces,
primarily around Transfer Mirror 2.
Since the walls are within a few mK of the sky temperature,
the un-corrected error signal after Fourier transformation
is thus of order
\begin{equation}
I_\nu = 0.003 ~ 
\times \frac{\partial B_\nu}{\partial T} ~
\times \delta T_{\rm inst} 
\label{syserr_amp_stray_2}
\end{equation}
or $10^3$~Jy~sr$^{-1}$.
As with the active optical elements,
the skymap software includes nuisance terms proportional 
to the variation in wall temperature
({\it cf} Eq. \ref{nuisance_eq})
to separate emission originating within the instrument
from true sky signals.
After correction,
all stray-light terms
contribute 
$\delta I_\nu < 10^{-3}$~Jy~sr$^{-1}$.

\begin{figure}[t]
\centerline{
\includegraphics[width=5.0in]{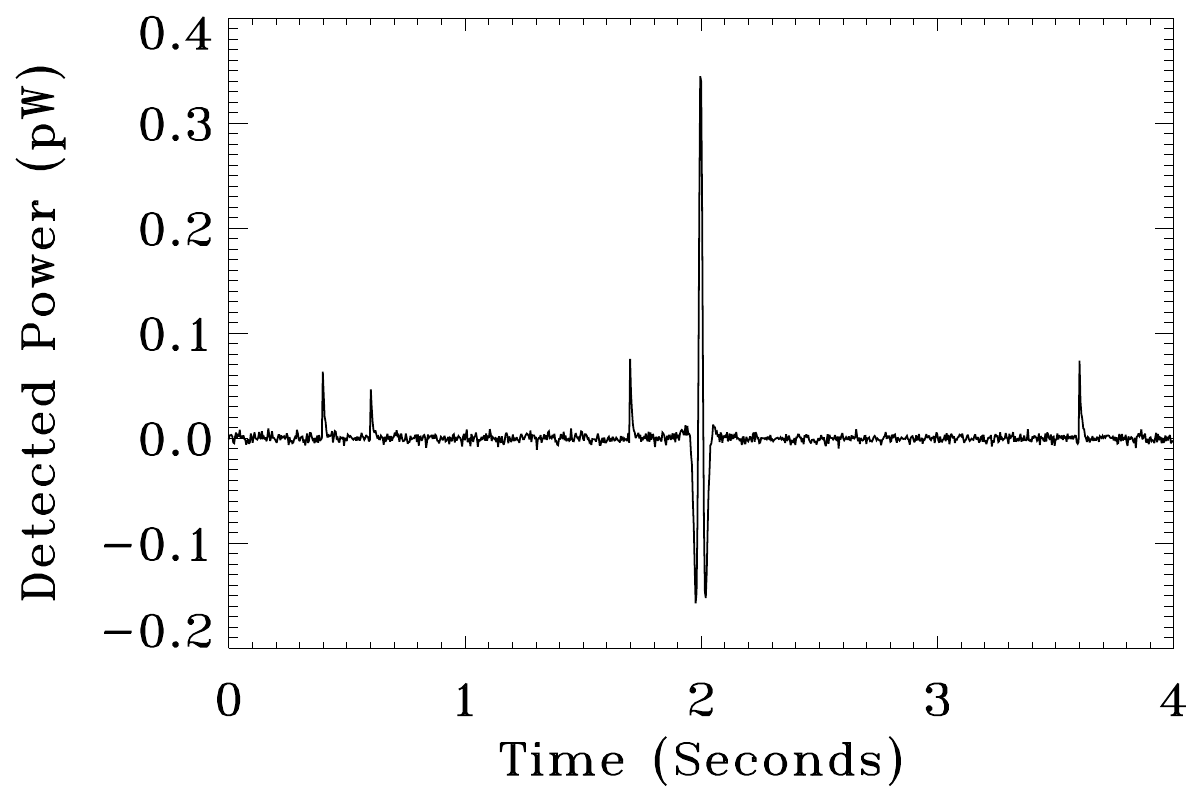}}
\caption{
Simulated interferogram including sky signals,
instrument noise, and random cosmic-ray hits.
Individual cosmic-ray hits to the detector absorber
are observed at high signal-to-noise ratio
and are co-added to create a template for cleaning.
}
\label{cosmic_ray_tod}
\end{figure}

Internal reflections can also contribute a systematic error signal.
The instrument calibration
and subsequent sensitivity to CMB spectral distortions
depends on in-flight observations of a full-aperture blackbody calibrator.
The calibrator consists of an array of absorbing cones
mounted on an thermally-conductive aluminum backplate,
with power reflection coefficient at normal incidence
$R < -65$~dB
\cite{kogut/fixsen:2019}.
The cone tip heights are staggered 
and the entire calibrator is tipped by 2\deg~
to mitigate standing waves within the instrument.
Rays originating from the detector
and reflecting from the calibrator
terminate elsewhere within the instrument.
Since the instrument and calibrator are maintained 
within a few mK of each other,
the uncorrected error signal
\begin{equation}
I_\nu = 10^{-6.5} 
\times \frac{\partial B_\nu}{\partial T} ~
\times \delta T_{\rm inst} 
\label{syserr_amp_cal}
\end{equation}
is below 1~Jy~sr$^{-1}$.
Since the reflected rays terminate
on components under active thermal control,
reflections are accounted for in the above-described nuisance parameters.

\subsubsection{Cosmic Rays}

Galactic cosmic-ray protons
in the energy range 0.1--1 GeV 
can impact the detector assembly to deposit energy
uncorrelated with the sky signal,
creating an additional additive systematic effect.
Two effects dominate.
Energy deposited in the detector frame
creates a stochastic background of thermal fluctuations,
sourcing corresponding fluctuations in the bolometer output signal.
This effect,
common to all bolometric detectors,
can be mitigated by tightly coupling the detector frame
to a well-controlled thermal bath.
PIXIE's single-crystal-silicon detector frames
include a 4 mm$^2$ pad of 4~$\mu$m thick electroplated gold
(residual resistivity ratio RRR $>$ 50)
on each corner of the frame.
Gold wires bonded to each pad
facilitate heat transport from the frame to the bath,
following the design from the
Hitomi Soft Xray Spectrometer microcalorimeters
\cite{kilbourne/etal:2018}.
The predicted thermal fluctuations for PIXIE
from cosmic ray hits to the detector frame is
$0.01~\mu$K~Hz$^{-1/2}$,
a factor of 30 below the thermal noise from
the adiabatic demagnetization refrigerator.

\begin{figure}[t]
\centerline{
\includegraphics[width=5.0in]{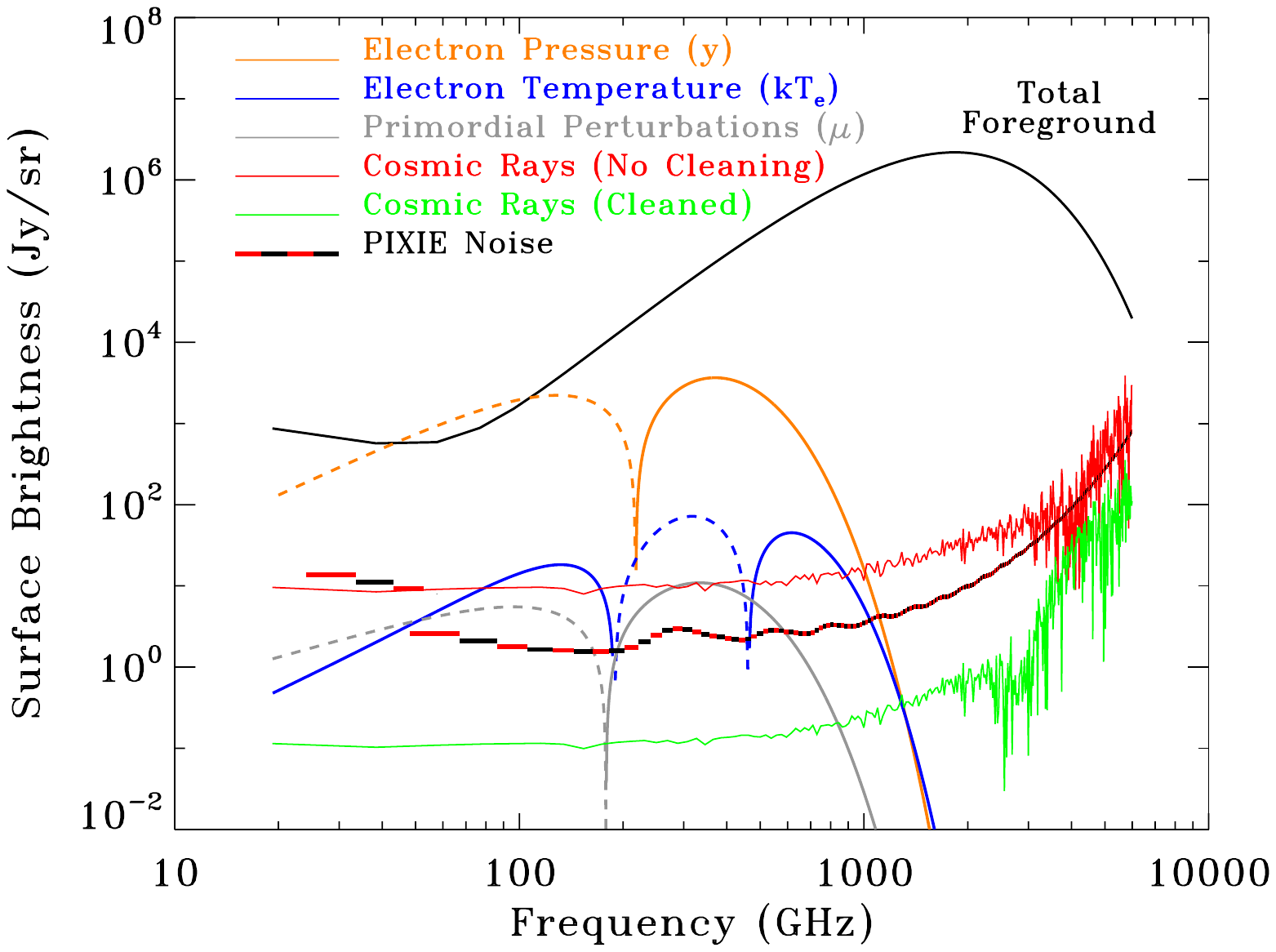}}
\caption{
Synthesized frequency spectra generated from a noiseless simulation
of cosmic-ray events over a 2-year mission
are compared to the instrument noise
and unpolarized sky signals.
Dashed lines show negative signals.
Cleaning the individual cosmic-ray events
with 1\% accuracy reduces the residual to negligible levels.
}
\label{cosmic_ray_cleaning}
\end{figure}

Cosmic ray hits directly on the bolometer absorbing structure
cannot be reduced by additional heat sinking,
but must be dealt with in the data processing pipeline.
Figure \ref{cosmic_ray_tod}
shows a simulation of cosmic-ray hits
to a single interferogram.
The Planck mission,
operating near solar minimum with correspondingly 
high rates of Galactic cosmic ray events,
saw hit rates of order 1--2~min$^{-1}$
directly to the spider-web absorber
\cite{planck_cosmic_ray_2014}.
The PIXIE absorber grid is larger
(12.7 $\times$ 12.7 mm with 11\% fill factor),
leading to an estimated hit rate of 1--2 hits per second.
A typical cosmic ray hit deposits 1.8 keV of energy
within the 1.4~$\mu$m thick silicon grid.
The signal-to-noise ratio 
is thus 15--20 at the peak of a single event,
increasing to 40 when integrated over the subsequent thermal decay.
A 2-year mission provides $8 \times 10^6$ cosmic ray hits
to be co-added for each detector
(5\% of the total time-ordered data samples),
providing an excellent basis for an empirical model
of the resulting thermal decay.

As with Planck,
individual cosmic ray events will be flagged and co-added
to produce a model for cleaning the time-ordered data.
Figure \ref{cosmic_ray_cleaning}
compares the effect of cosmic ray hits
before and after cleaning.
We generate noiseless simulations of time-ordered cosmic-ray events
for a full two-year mission,
Fourier transform each interferogram
to generate frequency spectra,
and sort the spectra by spin angle and sky pixel
to produce IQU data cubes.
If nothing were done to identify or clean cosmic-ray hits,
the resulting cosmic-ray noise in the synthesized spectra
would lie a factor of ten above the instrument noise\footnote{
Both the instrument noise and cosmic-ray spectra
are nearly independent of frequency in the Fourier-transformed spectra.
The rise in noise amplitude at frequencies above 1~THz
results from referring the detected noise to the sky signal,
which is attenuated at high frequencies by a set of low-pass filters.
}.
Cleaning the data
reduces the residual amplitude of individual cosmic-ray events
and modifies the time-ordered structure,
replacing the original sharp peak and subsequent decaying exponential
with
a smaller and smoother residual.
Here we model the residual as a damped sinusoid
with period twice the thermal time constant of the absorber,
and adjust the amplitude
to show a residual with 1\% of the power in the un-cleaned signal.
Cleaning at this level reduces the cosmic-ray noise
to negligible levels.

\begin{figure}[t]
\centerline{
\includegraphics[width=3.2in]{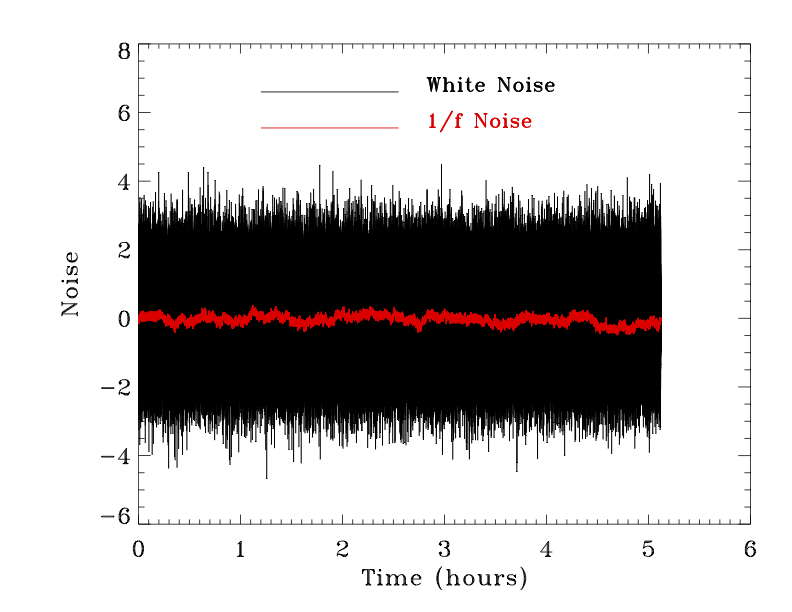}
\includegraphics[width=3.2in]{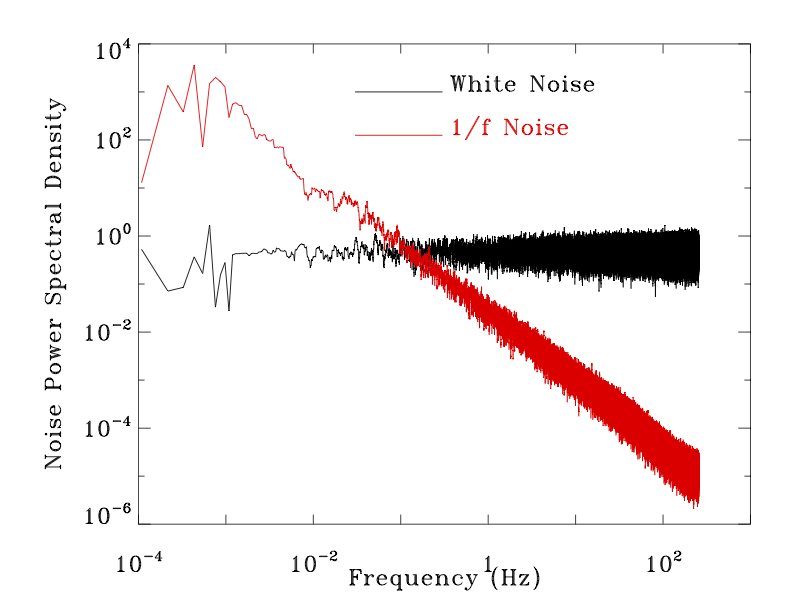}}
\caption{
Time ordered data (left) and power spectral density (right)
are shown
for simulated white and $1/f$ noise components \ 
over a single great-circle scan.
The $1/f$ knee frequency of 0.1 Hz
is fast compared to the 48 seconds required
for the instrument boresight 
to sweep across a single pixel,
but slow compared to the 4-second FTS phase delay stroke.
}
\label{1_over_f_example}
\end{figure}

\subsubsection{Correlated Noise
\label{one_over_f_section}}
Noise with non-zero correlation in the time domain --
often called $1/f$ noise for its frequency dependence --
is a common source of additive systematic error.
Figure \ref{1_over_f_example} shows an example.
Depending on the scan strategy, 
correlations between successive samples
lead to pixel-to-pixel correlation (striping) in the map domain.

\begin{figure}[t]
\centerline{
\includegraphics[width=6.5in]{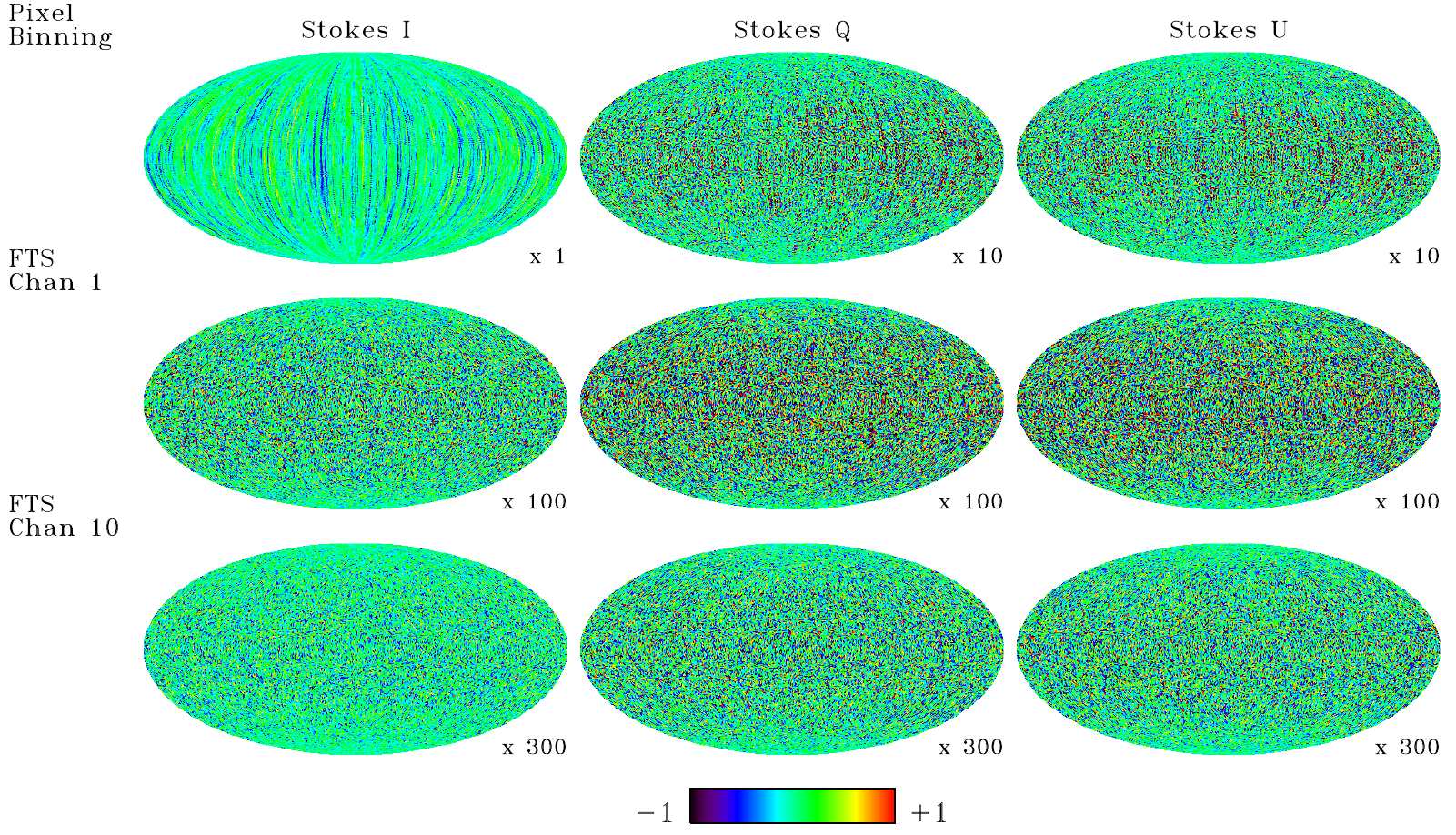}}
\caption{
Sky maps generated from simulated $1/f$ noise
are shown for the PIXIE scan pattern
(Mollweide projection in ecliptic coordinates).
Simply binning the time-ordered data by sky pixel
leads to  striping in the maps,
while the FTS
effectively whitens the $1/f$ component.
Maps have been scaled in amplitude as noted
to allow a common color stretch.
}
\label{1_over_f_maps}
\end{figure}

\begin{figure}[t]
\centerline{
\includegraphics[width=3.0in]{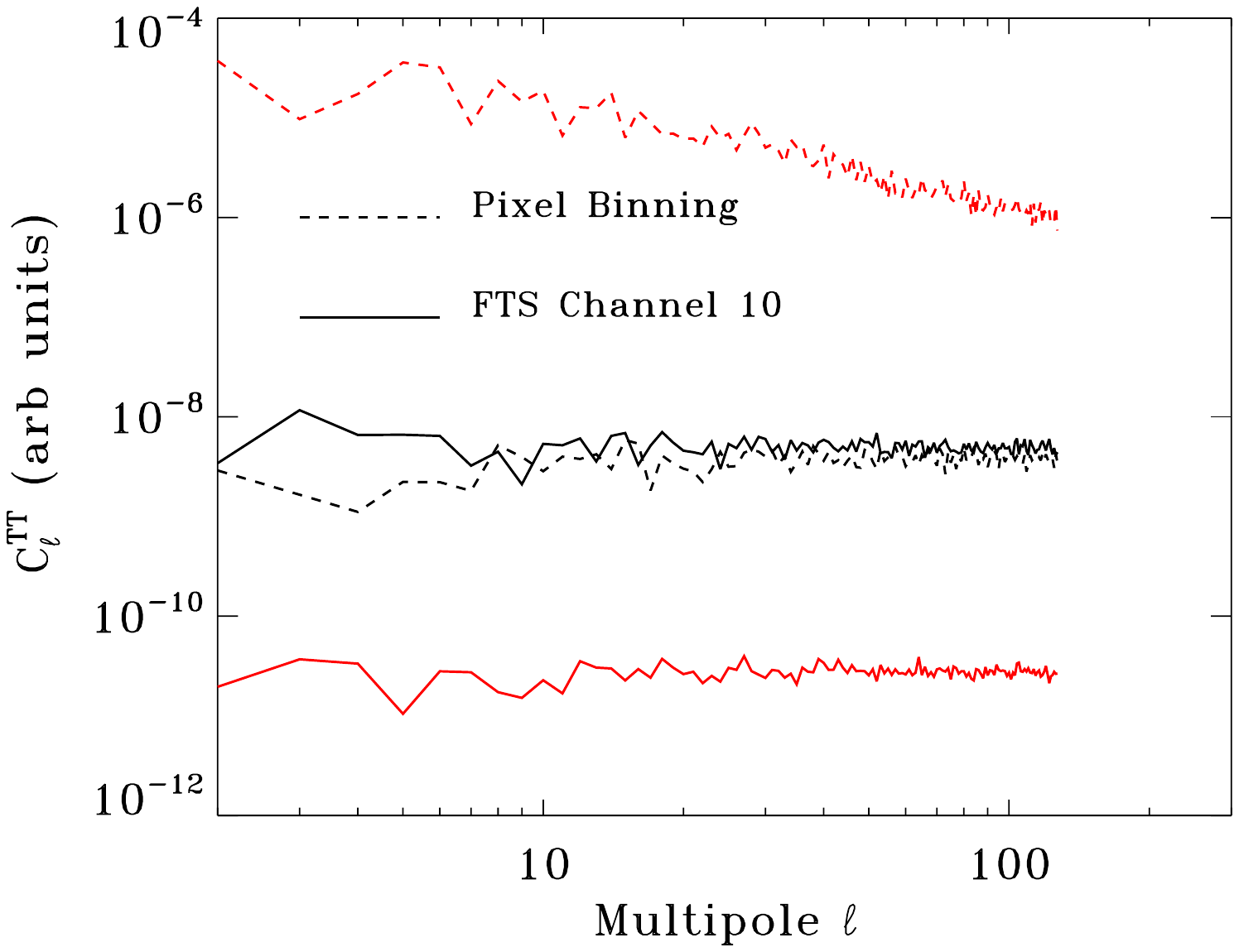}
\includegraphics[width=3.0in]{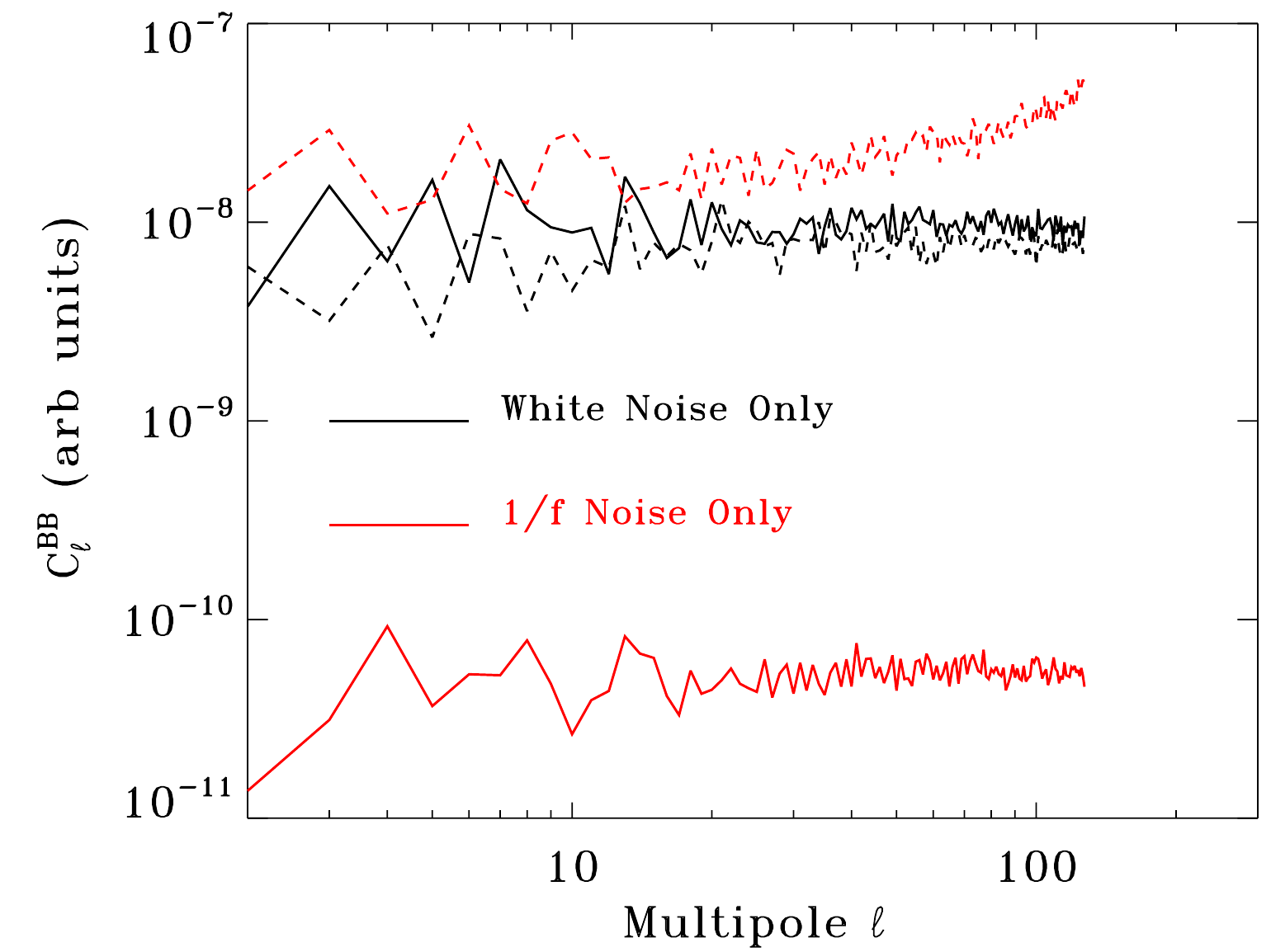}
}
\caption{
Angular power spectra for the maps shown in Figure \ref{1_over_f_maps}.
The left panel shows the temperature ($TT$) power spectra
while the right panel shows the B-mode ($BB$) power spectra.
Black lines in both planels represent the unit-amplitude white noise component,
while red lines represent $1/f$ noise with amplitude scaled to produce
knee frequency 0.1~Hz (see also Fig \ref{1_over_f_example}).
Dashed lines of either color show the angular power spectra
for simple pixel binning of the time-ordered data,
while solid lines show the angular power spectra
for the spatial map corresponding to a single synthesized frequency channel 
(channel 10, $\sim$270 GHz)
after Fourier transforming the time-ordered data.
The usual $\ell (\ell+1) / (2\pi)$ normalization
has been suppressed so that white noise shows as a horizontal line.
While the white-noise amplitudes are comparable
for both the pixel binning (dashed lines) and FTS maps (solid lines),
the FTS whitens and suppresses the $1/f$ component.
}
\label{1_over_f_cl}
\end{figure}

Figure \ref{1_over_f_maps}
shows 6-month sky maps generated from a
random noise simulation
consisting of $1/f$ noise with knee frequency 0.1~Hz.
As in $\S$\ref{periodic_section}, 
we compare sky maps
for simple pixel binning (without the FTS)
to the spatial/frequency sky cubes
produced by the FTS.
The spatial striping is evident
for pixel binning,
both for Stokes I and in polarization.
The PIXIE Fourier transform
effectively whitens $1/f$ noise.
Figure \ref{1_over_f_cl}
compares the angular power spectra
for the white noise and $1/f$ components
within the two mapping schemes.
The white-noise component appears at comparable amplitudes
regardless of the mapping scheme,
but the FTS provides orders-of-magnitude suppression
for the $1/f$ component
in both the polarized (BB) and unpolarized (TT) power spectra
as compared to simple pixel binning.
The FTS samples the $1/f$ noise at frequencies between
the 256~Hz detector sampling
and the 0.25 Hz mirror stroke.
The constant-velocity FTS mirror stroke
produces a linear relation between
the frequencies at which the data are sampled
and the optical frequencies after Fourier transformation,
$\nu_{\rm data} = v/c ~\nu_{\rm opt}$
where $v = 4$~mm~s$^{-1}$
is the phase velocity of the mirror stroke.
The projected $1/f$ noise is thus
largest in the lowest synthesized frequency channel
and decays as $\nu^{-1}$
for higher channels.

\subsection{Multiplicative Effects}

Multiplicative effects modulate the input signal difference.
The differential design and operation near null
reduce the amplitude of such effects,
but do not entirely eliminate them.
End-to-end simulations evaluate the response in the PIXIE data.

\subsubsection{Calibration Error}
\label{gain_error}

The readout electronics amplifies the signal from the detectors
and digitizes it for downlink telemetry.
The instrument calibration 
relies on observations of sources with known emission
to convert the downloaded detector output
from digitized telemetry units
to physical units of surface brightness.
Time-dependent variation in the resulting calibration,
if not corrected in the analysis,
can propagate through the Fourier transform
to produce systematic errors in the resulting sky spectra.

\begin{figure}[b]
\centerline{
\includegraphics[width=3.5in]{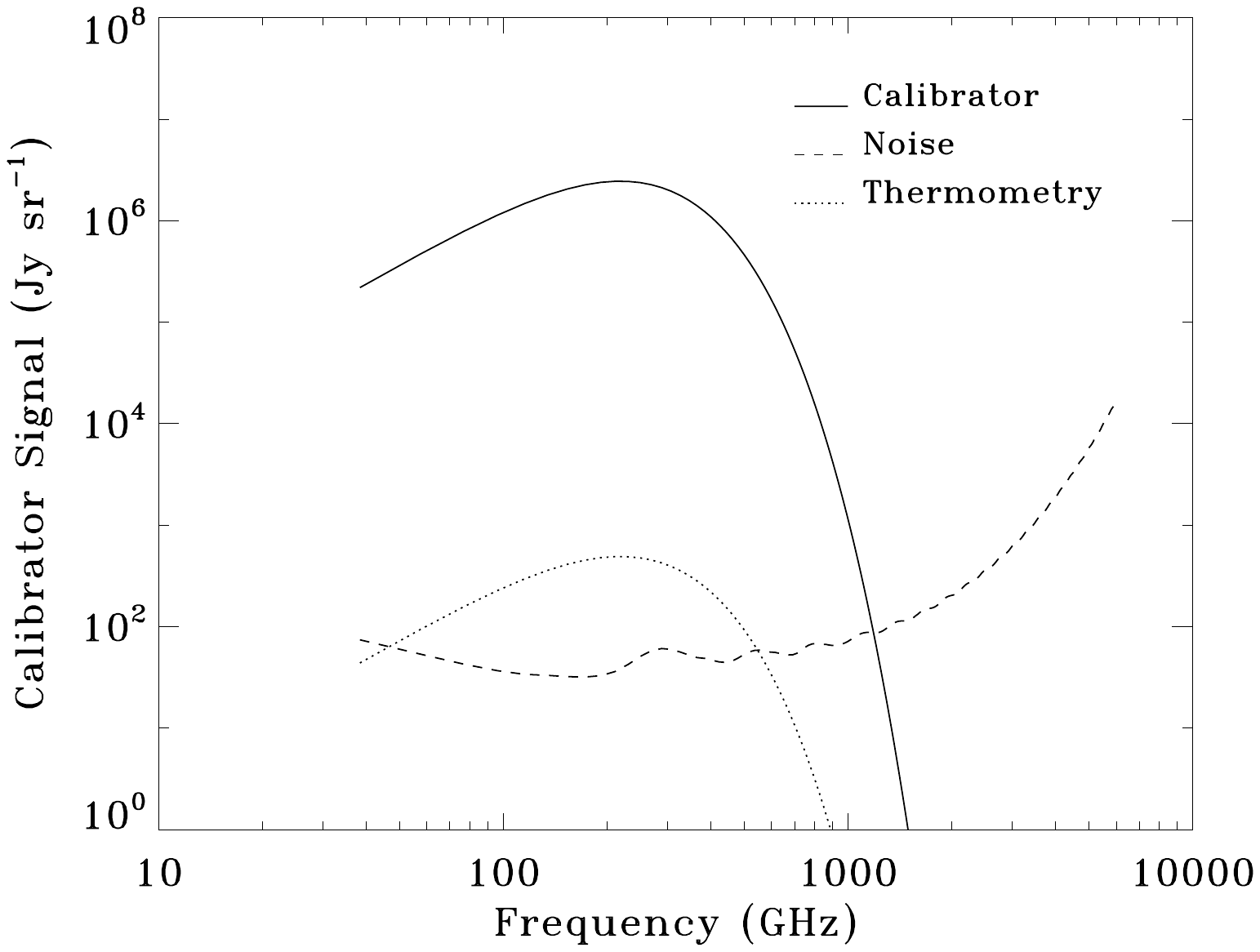}}
\caption{
The calibration signal 
from a 5 mK temperature change in the external calibrator
is large compared to 
instrument noise during the measurement
or the 1~$\mu$K differential thermometry precision.
Calibration drifts can be tracked to few-part-per-million precision
on time scales of a few hours or longer.
}
\label{xcal_signal}
\end{figure}

Instrument calibration 
compares data taken during two consecutive great-circle scans,
one with the calibrator blocking the A-side beam
and the second with the calibrator blocking the B-side beam.
Since input signals change sign
when observed in the A beam vs the B beam
(Eq. \ref{full_p_eq}),
the sky signal in each map pixel
cancels when summing the scan data for that pixel,
leaving only the difference in the calibrator signal.
The differential calibrator signal
averaged over the scan circle
provides an absolute blackbody reference
for signal calibration. 
A 5 mK temperature change in the external calibrator
produces a signal difference 
3--5 orders of magnitude 
above the instrument noise
or thermometry uncertainty
(Figure \ref{xcal_signal}),
allowing determination of the instrument calibration
to a few parts in $10^6$ every 10 hours throughout the mission
\cite{kogut/fixsen:2019}.

Calibration errors at this level produce negligible artifacts 
in the sky spectra.
After calibration, the data for a single detector may be written as
\begin{equation}
P_{Lx, \rm{cal}} = \left[ ~G_0(t) + \delta G(t) ~\right] 
~\left[
~\smallint \{ ~(E_{Ax}^2+E_{By}^2)+(E_{Ax}^2-E_{By}^2) \cos(z\omega /c) ~\}d\omega
~\right]   
\label{cal_equation}
\end{equation}
where
$G_0(t)$ is the true calibration (which may vary in time)
and
$\delta G(t)$ represents the error.
The constant term $E_x^2 + E_y^2$
is dominated by the CMB monopole
which does not vary across the sky;
anisotropies in the CMB or Galactic foregrounds
(although measured)
change the total power signal by less than 0.25\%
across the sky.
Gain errors 
$\delta G(t) \times (E_x^2 + E_y^2)$
in this term
are thus equivalent to additive systematic errors
discussed above.
With $\delta G / G \lsim 10^{-6}$,
the resulting error signal
has amplitude of a few $\mu$K in the time domain
prior to mapping,
and less than 1 nK after mapping
({\it cf} 
$\S$\ref{periodic_section})
 and 
$\S$\ref{one_over_f_section}.

\begin{figure}[t]
\centerline{
\includegraphics[width=2.1in]{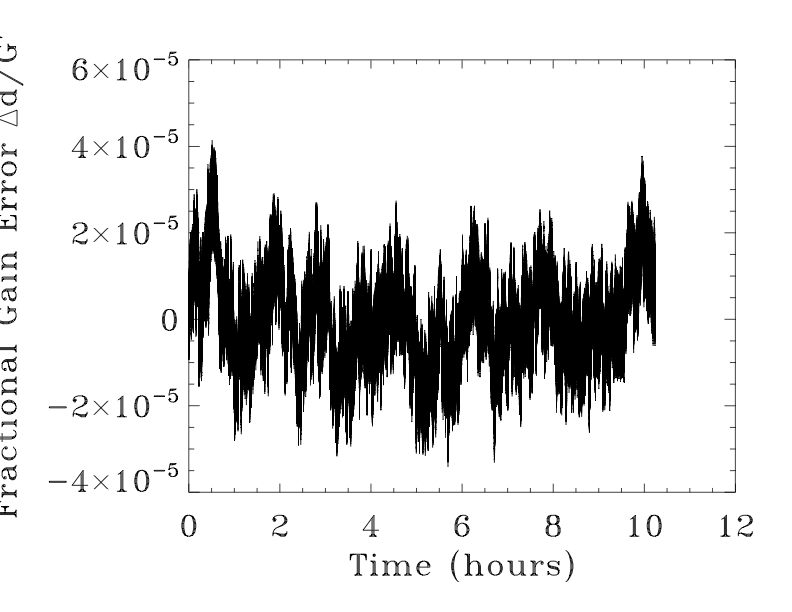}
\includegraphics[width=2.1in]{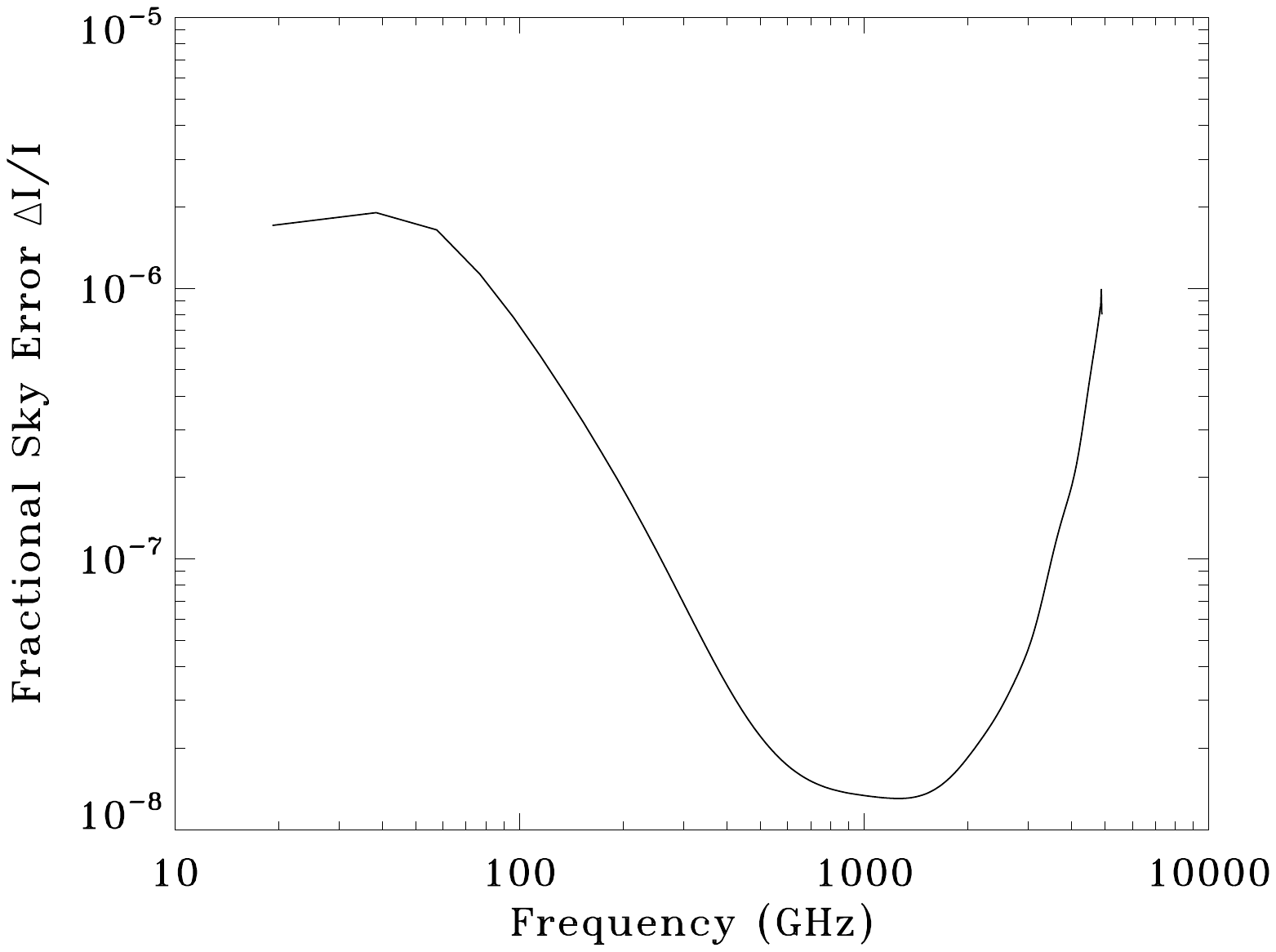}
\includegraphics[width=2.1in]{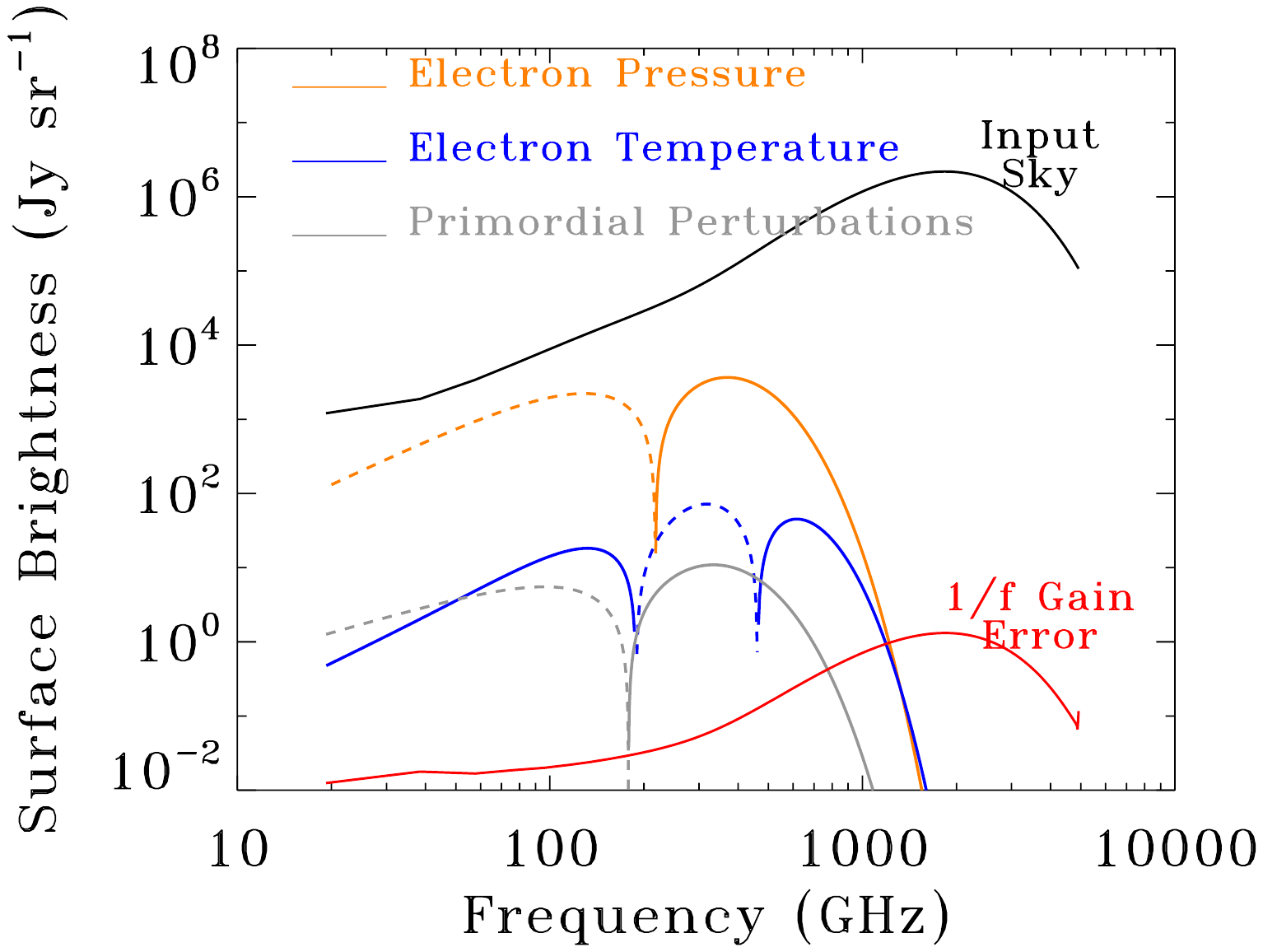}}
\caption{
(Left) Simulated $1/f$ gain variation with 1 mHz knee frequency.
(Center) Fractional error in the recovered sky spectrum
resulting from $1/f$ gain drift.
(Right) The systematic error in the recovered sky spectrum is small
compared to the cosmological signals.
}
\label{1_over_f_gain_fig}
\end{figure}

Gain errors in the Fourier transform are similarly small.
Fluctuations in the post-detection electronics
can produce $1/f$ variation in the calibration
as well as
variation locked to specific instrument time scales
(e.g. spin).
Figure \ref{1_over_f_gain_fig} shows an example 
of $1/f$ gain error.
We first generate the ideal interferogram $P(z)$
as a function of mirror position $z$
which would be produced by a typical unpolarized sky signal,
including median foreground contributions
as well as the 30~$\mu$K CMB anisotropy.
We then generate a $1/f$ time series $\delta G(t)$
with rms amplitude $10^{-5}$ and 1~mHz knee frequency
for a single 10-hour calibration period
to produce the time-ordered error signal
\begin{equation}
\delta P = \frac{\delta G(t)}{G_0} ~ P(z(t))
\label{gain_error_eq}
\end{equation}
in successive interferograms.
The Fourier transform of each successive interferogram
produces a series of sky spectra,
which we co-add to derive the
mean error in the calibrated sky spectra.
Note that the mirror reverses direction after every stroke,
effectively negating the contribution from gain drifts
on time scales longer than the mirror stroke.
As expected,
the effect of slow gain drifts is most pronounced
in the first few synthesized frequency channels.
A fractional gain error  $\delta G(t)/G = 10^{-5}$
produces a comparable fractional error 
$\Delta I/I$
in the first few channels of the sky spectra,
and falls rapidly for higher frequency channels.
The resulting error $\Delta I_\nu$ in the sky spectra
is below 1 Jy sr$^{-1}$ at all frequencies
and is negligible compared to the instrument noise
or cosmological signals.
 
\begin{figure}[b]
\centerline{
\includegraphics[width=4.0in]{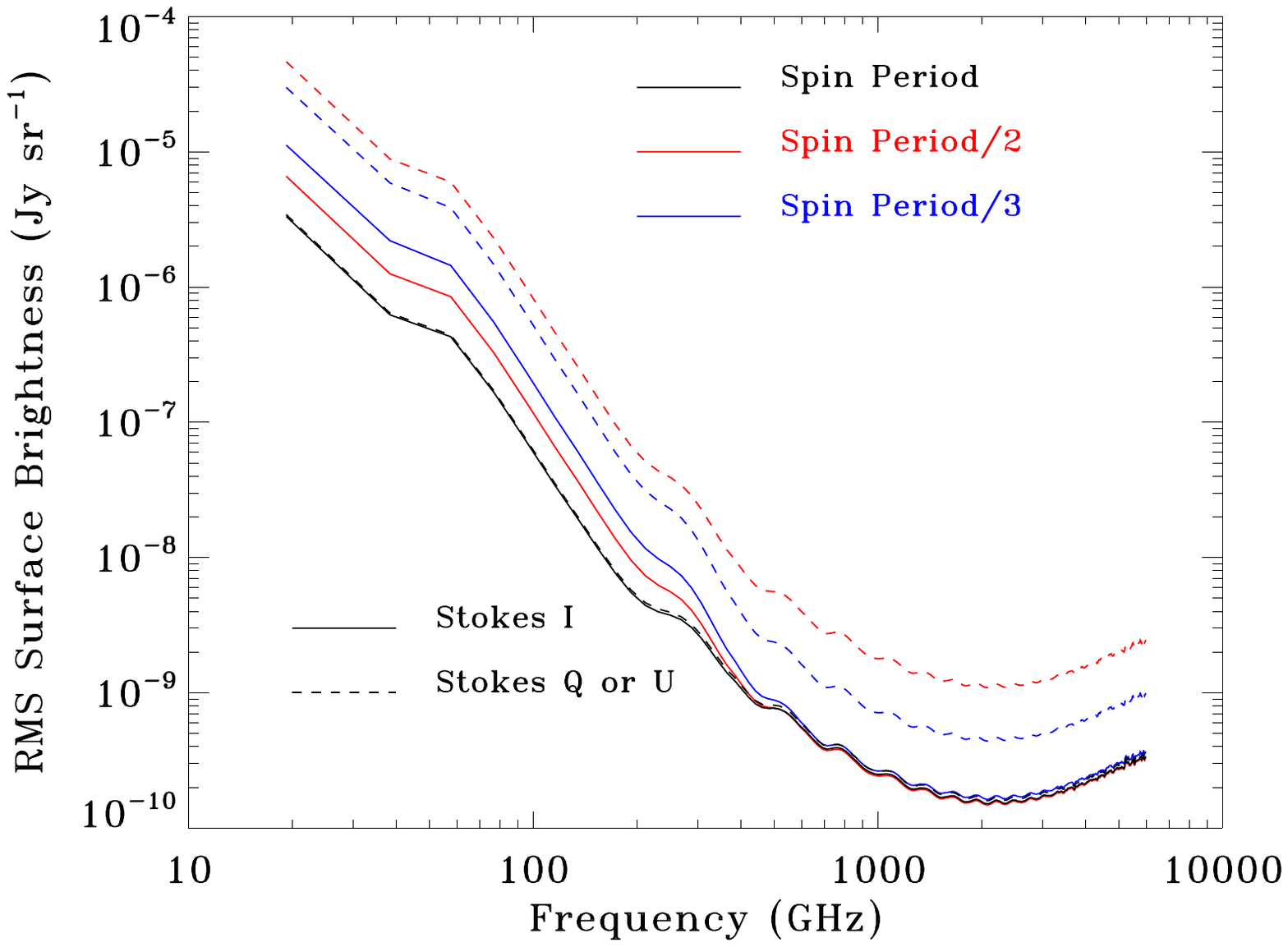}}
\caption{
Thermal variation of the instrument electronics
can modulate the post-detection gain
to create striping in the maps.
The figure shows the RMS amplitude of the striping
for gain modulation 
with amplitude $\delta G/G = 10^{-4}$ 
at selected harmonics of the spin period.
Solid lines show the spectra for unpolarized emission (Stokes I)
while dashed lines show the polarized spectra (Stokes Q or U).
}
\label{spin_gain_spectra}
\end{figure}

Gain modulation at harmonics of the spin period 
can be sourced by corresponding variation 
in the instrument temperature or power system.  
The 48-second spin period 
with axis perpendicular to the sun line
produces a ``barbecue roll'' illumination
of the observatory exterior surface
(Figure \ref{pixie_scan}).
The 75-second barbecue roll
for the Cosmic Background Explorer
produced synchronous variation 
in the instrument electronics temperature
below 10 mK
\cite{dmr_syserr_1992}.
Electronics components typically show
thermal gain variation at levels of a few percent per K.
The faster PIXIE spin would thus be expected
to produce spin-locked gain variation
at levels $\delta G/G \approx 10^{-4}$ or less.
As with $1/f$ gain modulation,
the primary effect of spin-locked modulation
is on the detector total power,
generating spatial striping similar to the 
additive signals in Figure \ref{spin_sine_maps}.
Figure \ref{spin_gain_spectra}
shows the rms amplitude of the 
spin-locked signals
for selected harmonics of the spin period.
Gain modulation of the detector total power
produces negligible artifacts in either total intensity or polarization.
Variation in the detected power as a function of mirror position
is less than 1\% of the total power:
spin-locked gain modulation 
of the interferogram produced by anisotropic sky signals
is even smaller.

Errors in the calibrator thermometry can also source systematic errors.
Each of the thermometers within the external calibrator
are read out once per second
with read noise 1~$\mu$K,
which integrates down to few-nK precision
during a single 10-hour calibration cycle,
The absolute thermometry scale is set in flight
to accuracy 10~$\mu$K
using observations of Galactic emission lines
and the Wien displacement law
for the measured blackbody spectra
\cite{firas_syserr_1994,
kogut/fixsen:2019}.
Since the sky spectra are referenced against to the blackbody calibrator,
errors in the absolute thermometry
affect the calculated monopole temperature $T_0$
but cancel to first order for spectral distortions,
resulting in systematic error 
$\delta I_\nu < 5 \times 10^{-3}~$~Jy~sr$^{-1}$
for the derived spectral distortions.

\subsubsection{Bandpass Error}

Comparison of signals between different detectors
can lead to systematic error
if different detectors have different frequency response.
Sharp spectral features such as line emission
are particularly problematic.
Measurements of polarization
by the Planck mission
required careful correction 
for the differential bandpass of individual detectors
\cite{
planck_hfi_bandpass_2013,
planck_hfi_processing_2016,
planck_hfi_processing_2020,
beyond_planck_bandpass_2022}.
PIXIE does not use bandpass filters to define individual spectral channels,
but instead relies on Fourier transformation
of the measured interferograms
to derive the spectra
within synthesized frequency channels.
An ideal FTS
allows a single detector
to produce data at a large number of
well-characterized frequency channels,
with
the channel width (and hence channel center frequencies)
determined by the maximum mirror throw,
the number of channels 
(and hence the highest synthesized frequency)
set by the number of detector samples within an interferogram,
and the channel-to-channel covariance
determined by the interferogram apodization
\cite{kogut/fixsen:2019}.
For convenience,
the PIXIE channel widths are commensurate with the
CO J=1-0 line at 115.3 GHz within the Galaxy,
$\Delta \nu  = \nu_{\rm CO} / N$
so that every $N^{\rm th}$ channel
is centered on a Galactic CO line.
The total optical passband 
is limited by scattering filters on the beam-forming optics.
Both the scattering filters
and the mirror sampling
are common to all four detectors.
Frequency dependence 
in the individual detector absorption coefficients
or the optical transmission from the detector to the sky
can be identified and characterized
by detector-to-detector comparison
of the blackbody calibrator
($\S$4.5).

\begin{figure}[b]
\begin{center}
\includegraphics[width=3.5in]{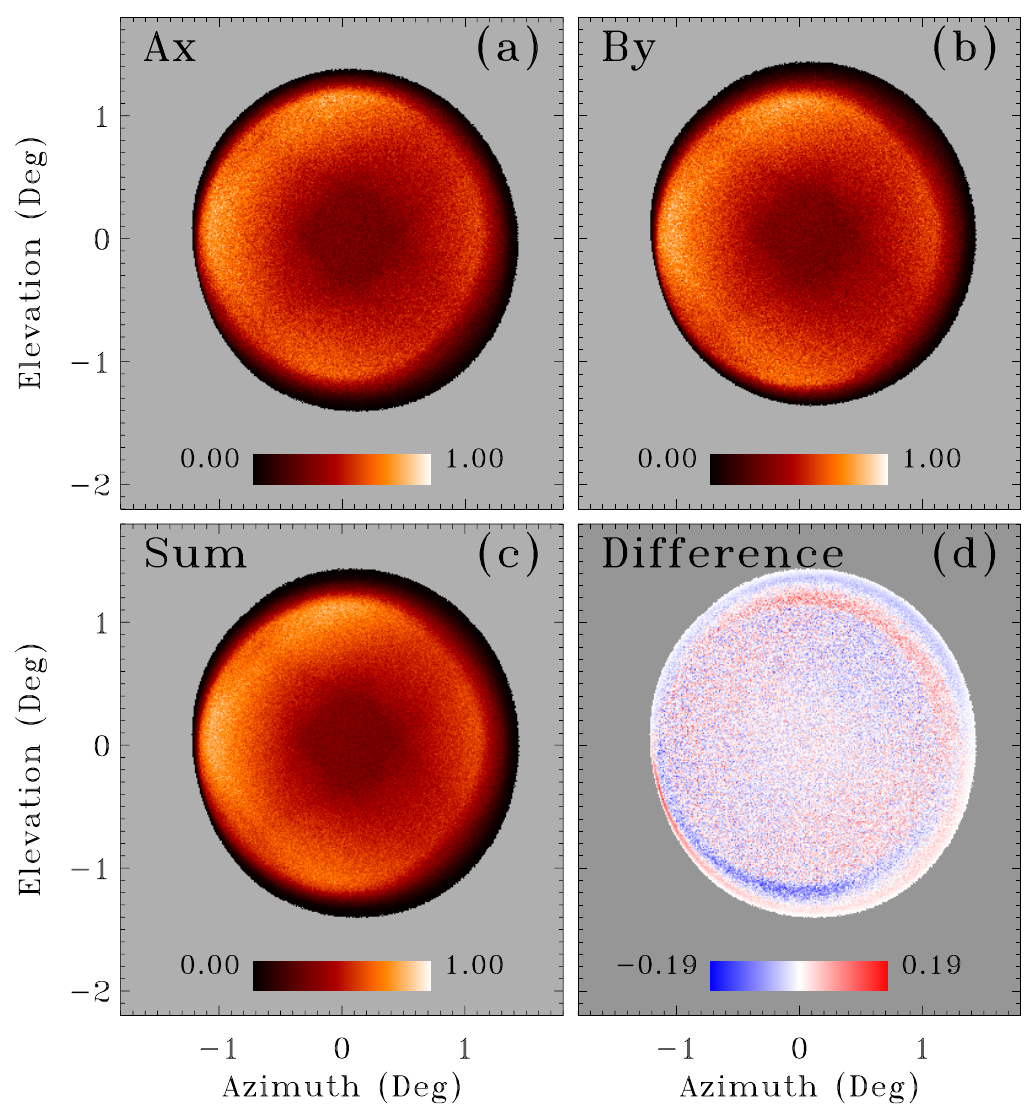}
\end{center}
\caption{ 
Monte Carlo realization of the beam patterns for the
$P_{Lx}$ detector,
generated from a ray trace of $10^{11}$ rays
from the detector through the optics to the sky.
(a) $\hat{x}$ polarization in the A-side beam.
(a) $\hat{y}$ polarization in the B-side beam.
(c) $(A_x + B_y)/2$ sum.
(d) $(A_x - B_y)/2$ difference.
Plots (a) and (b) are normalized to unity at the peak.
Tolerance errors in the individual optical elements
create a small offset in beam position
(see text).
\label{beam_maps}
}
\end{figure}

\subsubsection{Beam Errors}
\label{beam_section}

An extensive literature analyzes
potential systematic error and mitigation strategies
related to beam shape
\cite{
hu/etal:2003,
odea/etal:2007,
rosset/etal:2007,
shimon/etal:2008}.
This is of particular concern for polarization measurements,
which commonly rely on differencing the signal
between two beams on the sky.
Differential beam shape can then
produce leakage between polarization
and the much brighter unpolarized signals.

PIXIE's differential design and mission operations
mitigate beam errors.
Each detector is intrinsically sensitive to the
difference in polarization 
seen by two co-pointed beams
(Eq. \ref{full_p_eq})
so that systematic errors 
coupling unpolarized temperature gradients
to a false polarized signal
cancel to first order for any individual detector
and to second order when comparing detectors
\cite{pixie_4_port_2019}.
The instrument is symmetric 
about the midline between the two beams
(Fig. \ref{symmetry_cartoon}),
so that the $\hat{x}$ polarization from one beam
is the mirror reflection
of the 
$\hat{y}$ polarization from the other beam.
Temperature-polarization mixing thus depends on the
anti-symmetric component
of the difference between the A and B beams on the sky.

\begin{figure}[t]
\begin{center}
\includegraphics[width=4.5in]{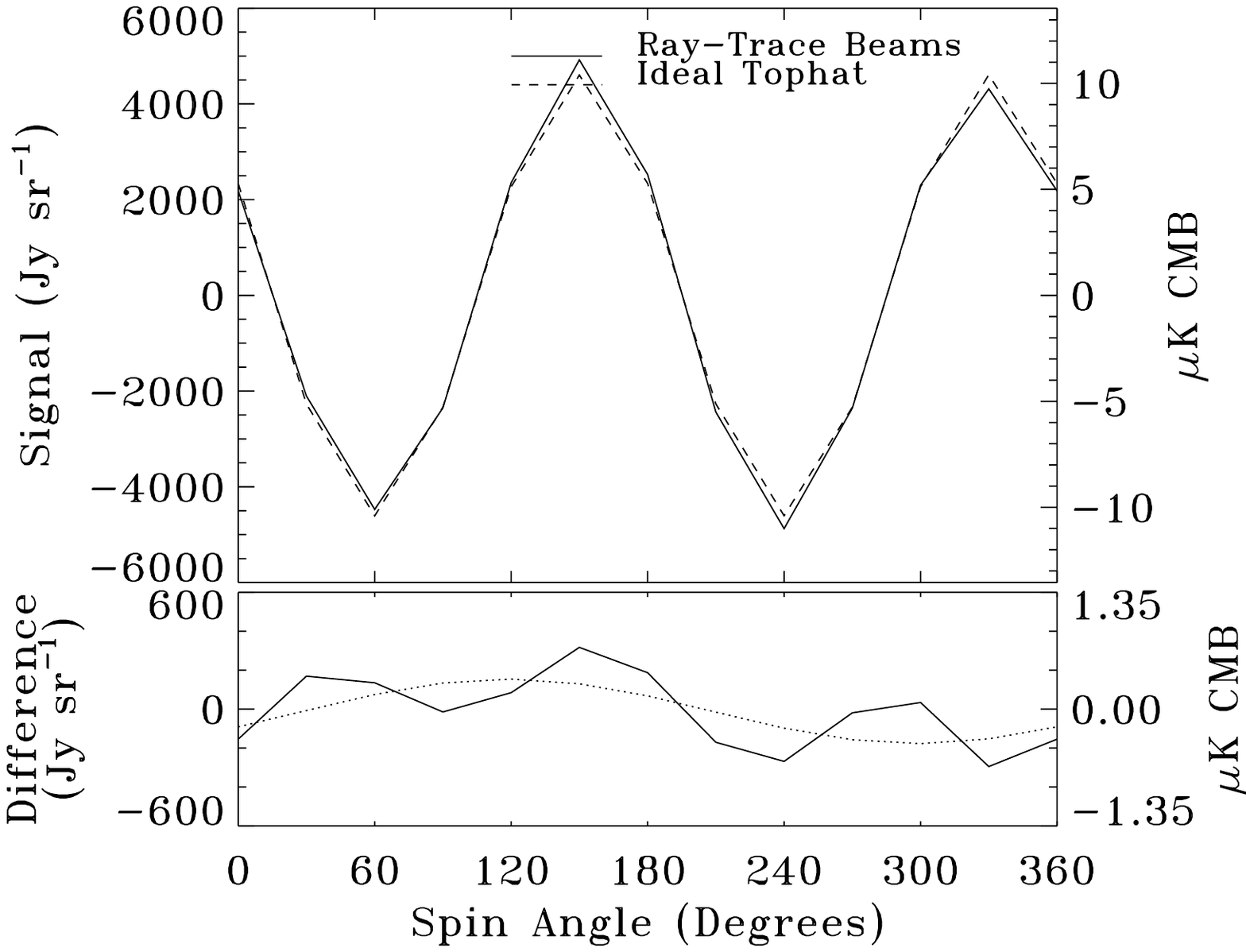}
\end{center}
\caption{ 
Comparison of the spin-modulated signal for a single PIXIE detector
between ideal tophat beams
and the  Monte Carlo ray-trace beam patterns
(Fig. \ref{beam_maps})
for a single mid-latitude pixel 
at 270 GHz.
The difference signal is dominated by a dipole term
(dashed line in bottom panel)
which does not couple to polarization.
\label{beam_comparison}
}
\end{figure}

PIXIE's multi-moded light-bucket optics
produce a circular tophat beam for each detector.
Since there is only one spatial pixel,
centered on the optical axis,
beam effects such as coma and aberration are minimized.
In this, PIXIE differs
from most CMB imaging instruments,
which employ kilo-pixel arrays of single-moded optics
to generate an array of Gaussian beams on the sky.
Figure \ref{beam_maps} shows simulations of the PIXIE beams
for the $\hat{x}$ detector on the left side of the instrument
($L_x$; {\it cf} Eq. \ref{full_p_eq}).
We use a Monte Carlo ray-trace code to propagate $10^{11}$ random rays
from the detector
through the full PIXIE optics to the sky.
Deviations in the position or orientation
of the concentrator and mirrors
will affect the beam pattern.
To account for such tolerance effects,
each Monte Carlo realization
independently perturbs each optical element
in both position and orientation
by an amount drawn from a Gaussian distribution
whose width is set by typical machining/assembly tolerances
of 0.05 mm.
The dominant effect of such tolerance errors
is an anti-symmetric response in the beam difference.

The anti-symmetric component of the A-B beam difference
couples to unpolarized gradients on the sky
to source systematic error.
The spacecraft spin allows identification and separation of
symmetric vs anti-symmetric signals.
The 48-second spin period is rapid compared
to the great-circle scan or annual orbital precession,
allowing the data within each pixel on the sky 
to be decomposed into $m$ moments of the spin angle.
True sky polarization is modulated
at twice the spacecraft spin frequency ($m=2$),
while
any anti-symmetric contribution
can only appear at odd harmonics of the spin.
Figure \ref{beam_comparison}
shows the resulting spin modulated response
for a typical mid-latitude pixel
(Galactic coordinates $[l, b] = [-120\deg, -50\deg]$)
observed at 270 GHz.
We compare the response
of the ray-trace beam realization
(including tolerance errors)
to the response
from ideal circular tophat beams 
of identical solid angle.
Both responses show the expected $m=2$ modulation
at twice the spin frequency.
The difference between the ideal and realistic beams
is dominated by a dipole ($m=1$) 
which does not couple to the fitted sky polarization.
Additional simulations of the PIXIE beams
quantify the response to beam errors.
For a single detector,
the beam response at $m=2$ is suppressed by 35~dB
relative to the monopole $m=0$ power.
This falls to -60~dB for the double-difference
when comparing different detectors.
Assuming unpolarized gradients of order 15~$\mu$K/deg
from from the CMB dipole and Galactic foregrounds,
the corresponding systematic response in polarization
is below 10 nK for a single detector
and below 0.1 nK for detector pairs
\cite{pixie_4_port_2019}.

Angular offsets between the beam centroids and the spacecraft spin axis
will also create a spin-modulated systematic error.
Each detector is sensitive to the difference between
the power in orthogonal linear polarizations
received from the two beams
(Stokes Q in instrument coordinates).
Rotation of the instrument about the beam axis
rotates the instrument coordinate system
relative to the sky
to allow separation of Stokes $Q$ and $U$ parameters on the sky,
\begin{eqnarray}
S(\nu)_{Lx} &=& \frac{1}{4} 
	\left[ ~I(\nu)_A - I(\nu)_B 
	+ Q(\nu)_{\rm sky} \cos 2\gamma + U(\nu)_{\rm sky} \sin 2\gamma ~ \right] 
	\nonumber \\
S(\nu)_{Ly} &=& \frac{1}{4}
	\left[ ~I(\nu)_A - I(\nu)_B 
	- Q(\nu)_{\rm sky} \cos 2\gamma - U(\nu)_{\rm sky} \sin 2\gamma ~\right]
	\nonumber \\
S(\nu)_{Rx} &=& \frac{1}{4} 
	\left[ ~I(\nu)_B - I(\nu)_A 
	+ Q(\nu)_{\rm sky} \cos 2\gamma + U(\nu)_{\rm sky} \sin 2\gamma ~ \right] 
	\nonumber \\
S(\nu)_{Ly} &=& \frac{1}{4}
	\left[ ~I(\nu)_B - I(\nu)_A 
	- Q(\nu)_{\rm sky} \cos 2\gamma - U(\nu)_{\rm sky} \sin 2\gamma ~\right]~,
\label{diff_spectra_eq}
\end{eqnarray}
where
$S(\nu)$ is the Fourier-transformed spectrum from each detector,
$I = \langle E_x^2 + E_y^2 \rangle$,
$Q = \langle E_x^2 - E_y^2 \rangle$, 
and
$U = 2 {\rm Re} \langle E_x E_y \rangle$
are the Stokes polarization parameters
and
$\gamma$ is the spin angle,
defined relative to meridians of ecliptic longitude.
If the beam centroids are not aligned with the spacecraft spin axis,
the beams trace circles on the sky centered on the spin axis.
The resulting sampling of different locations on the sky
will couple to spatial gradients on the sky
to source a spin-dependent systematic error.
Sky gradients
on scales larger than the 2.65\deg~tophat beam
are dominated by the CMB dipole,
sourcing gradients of typical scale $15~\mu$K/deg.
Galactic and zodiacal foregrounds contribute additional gradients.
Large-scale gradients combine with the spacecraft spin
to create a dipolar response with respect to the spin angle
(spin moment $m=1$).

\begin{figure}[t]
\centerline{
\includegraphics[width=5.0in]{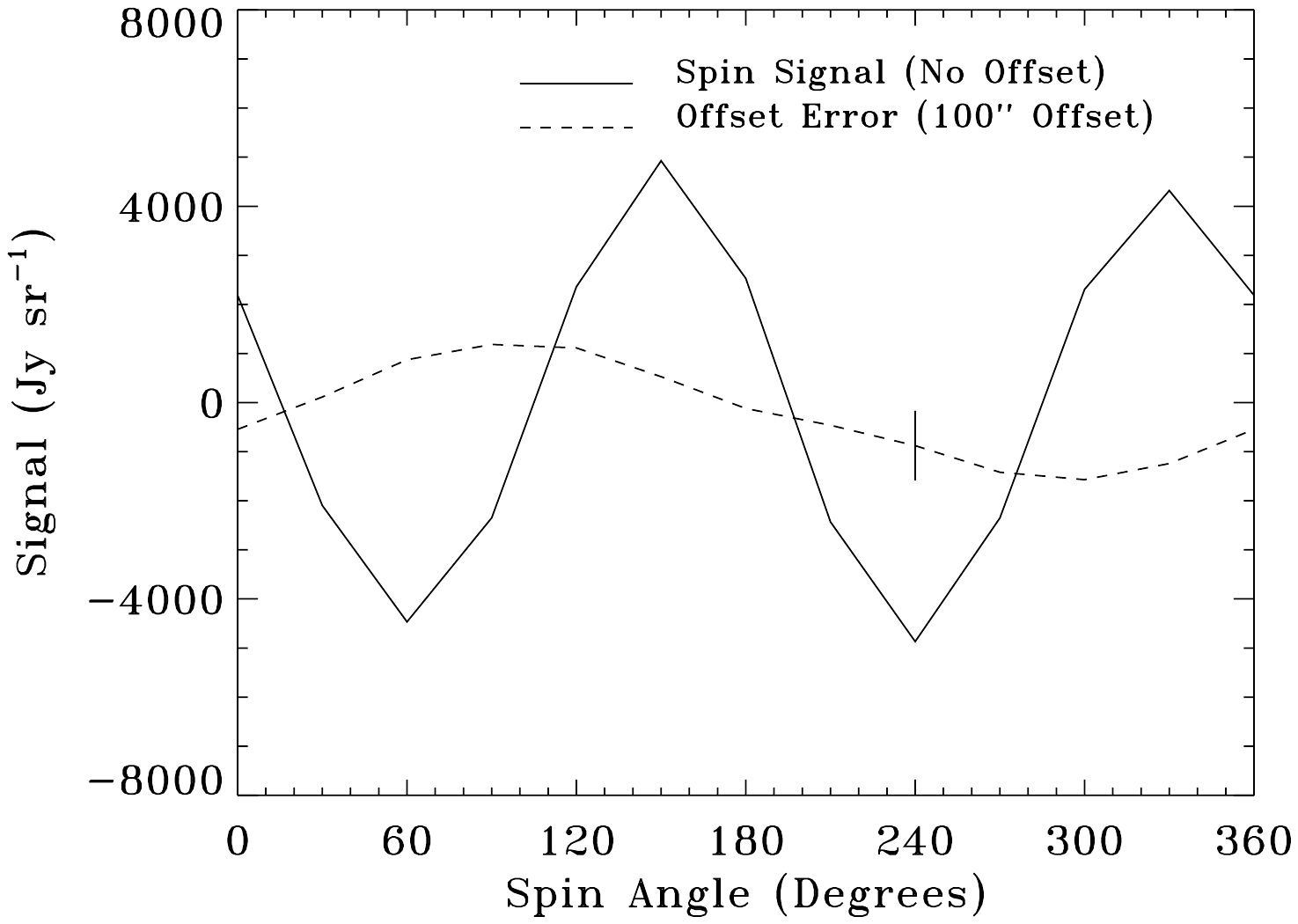}}
\caption{
The systematic error signal from a beam offset of 100\asec
is shown for polarized observations 
of a single
mid-latitude pixel 
($l$ = -120\deg, $b$ = -50\deg)
at a single synthesized frequency of 270 GHz
(the peak of the CMB $\partial B / \partial T$ anisotropy spectrum).
The solid vertical bar represents 
the noise at each spin angle.
The true sky polarization (solid line)
shows the expected $m=2$ dependence on spin angle,
while the systematic error (dashed line)
is dominated by a dipole ($m=1$) response.
}
\label{beam_offset_vs_angle}
\end{figure}

We evaluate the systematic error from fixed beam offsets
using simulations.
We first generate sky maps in each of the Stokes IQU parameters
using a superposition of
the CMB dipole, unpolarized anisotropy, and E-mode polarization,
to which we add the
Planck model of synchrotron and dust emission.
We then use  Eq. \ref{diff_spectra_eq} to generate the signal on a single detector
at each of 12 spin angles $\gamma$ uniformly spaced from 0 to $2\pi$.
At each spin angle,
we offset the A and B beams,
independently convolve the IQU sky maps
with the A and B beams,
then combine the beam-convolved maps
to produce the detected signal.
We model the beams
using a ray-trace realization including tolerance errors
(Fig. \ref{beam_maps}).

Figure \ref{beam_offset_vs_angle}
shows the systematic error signal 
resulting from a 100\asec offset between the A and B beams
observed
for a typical mid-latitude sky pixel
within a single synthesized frequency channel at 270 GHz.
With both beams open to the sky,
the true sky signal (solid line)
shows the expected $m=2$ dependence on the spin angle.
Unpolarized gradients source a smaller $m=1$ dipole (dashed line),
which can be compared to the measured sky gradient
to determine and correct the beam offset.

\begin{figure}[t]
\centerline{
\includegraphics[width=5.0in]{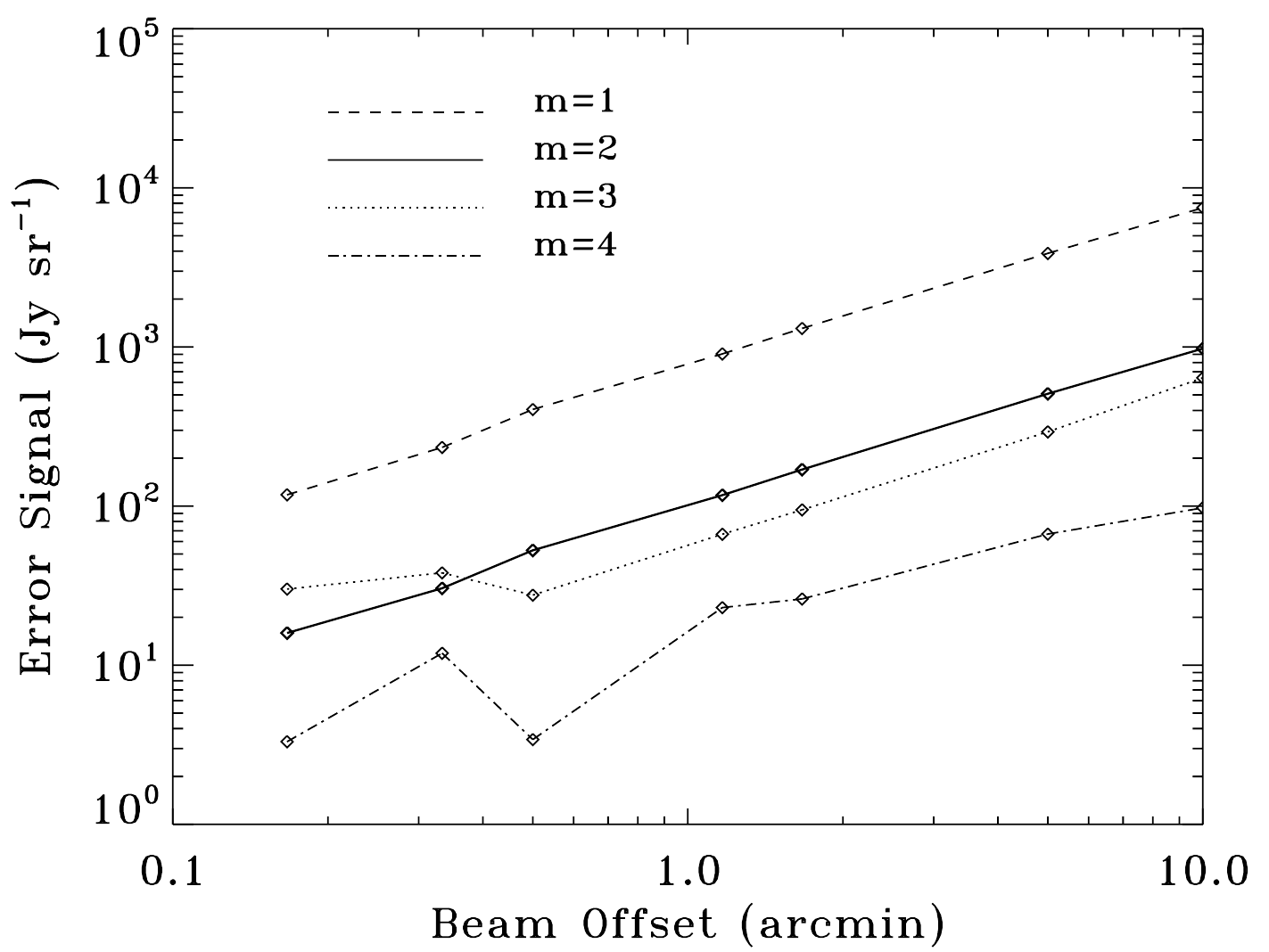}}
\caption{
The amplitude of systematic errors from beam offsets
depends on the distance 
by which the beam centroid differs from the spacecraft spin axis.
The resulting  differential beam patterns
couple to unpolarized gradients on the sky
to produce an anti-symmetric signal
dominated by the $m=1$ spin dipole.
}
\label{beam_offset_vs_m}
\end{figure}

The 48-second spin period
is commensurate with the 4-second mirror stroke
so that
a full rotation of the instrument relative to the sky
encompasses
12 mirror strokes uniformly spaced in spin angle.
Tagging each IFG within a pixel by the spin angle
enables a Fourier decomposition 
in harmonics $m$ of the spin frequency
to order $m=6$.
The uniform sampling in $\gamma$ within each pixel
minimizes covariance between the orders $m$.
Figure \ref{beam_offset_vs_m} 
shows the amplitude of the systematic error from beam offsets
as a function of spin moment $m$
(again evaluated at a single 270 GHz PIXIE channel).
As with errors due to internal beam structure
\cite{pixie_4_port_2019},
the systematic error at the cosmologically-relevant $m=2$
is suppressed by 35 dB relative to the dominant $m=1$ effect.
The nominal noise of 75 Jy~sr$^{-1}$ 
for observations of a single pixel at a single frequency channel
allow determination of the beam offsets to accuracy 6\asec,
reducing the residual $m=2$ error to negligible levels
(0.1~nK at 270~GHz).
In practice,
the mapping pipeline will include 4 additional free parameters
to describe the offsets of the A and B beams,
using the full multi-pixel, multi-frequency data set
to fit the beam centroids to few-arcsecond accuracy.

Measurements of spectral distortions
are much less sensitive to beam errors.
With the calibrator deployed to block one beam,
information on the unpolarized intensity
is contained within the $m=0$ spin mode
(Eq. \ref{diff_spectra_eq}).
Regardless of the actual beam shape,
the spin-averaged $m=0$ beam
is azimuthally symmetric.
Deviations from azimuthal symmetry in the indivudal beam profiles
couple to higher-order spin terms $m > 0$
and do not couple to spectral distortions
($\S$4.4).

\subsubsection{MTM Position Errors}
\label{MTM_jitter_section}

The Fourier transform from the measured interferograms
to the synthesized frequency spectra
assumes that the phase-delay mechanism
is in the correct position for each detector sample.
Differences between the nominal and actual positions
modulate the synthesized spectra
to create both random (noise) terms
and coherent systematic terms.
Nagler et al. (2015)
\cite{nagler/etal:2015} 
discuss the simplest case of a fixed phase offset,
either from a metrology error
or a timing error such that the detector samples
do not line up onto zero path difference (ZPD).
A fixed position offset $\Delta z$ in the IFG sampling
produces errors in both the real and imaginary parts
of the synthesized spectra,
\begin{eqnarray}
\Delta S_\nu^{\rm Re} &=& S_\nu \left( 2\pi (\nu \Delta z / c)^2 \right) \nonumber \\
\Delta S_\nu^{\rm Im} &=& S_\nu \left( 2\pi (\nu \Delta z / c) \right) ~,
\label{nagler_eq}
\end{eqnarray}
modulating to the incident spectrum $S_\nu$
by terms proportional to powers of the position offset $\Delta z$.
Since the imaginary Fourier transform
depends on $\sin(z)$ while the real part depends on $\cos(z)$,
for small offsets $\Delta z$
the effect appears at first order in $\Delta z$
for the imaginary spectra
and at second order $(\Delta z)^2$ for the real spectra.
For position errors $\Delta z$ less than the 7~$\mu$m mirror step size,
the effect in the real spectrum
is suppressed 
by 2--4 orders of magnitude
relative to the imaginary spectrum
over the PIXIE frequency range. 
ZPD errors may thus be removed
by fitting a single phase term $\Delta z$
to minimize the imaginary part of the Fourier transform.
A single phase correction
fit to the full mission data
removes a single degree of freedom
from over $10^7$ discrete IFGs,
reducing the error to levels 
$\delta S_\nu^{\rm Re} < 10^{-6}$~Jy~sr$^{-1}$
with negligible increase in noise.

Random errors in the mirror position
produce a related effect.
The interferogram amplitude
at the $i^{\it th}$ position sample
may be represented by the cosine transform
\begin{equation}
P_i = \int \left( S(\nu) \cos(2\pi \nu / c ~ z_i \right) d\nu
\label{cos_eq_1}
\end{equation}
up to normalizing constant
(Eq. \ref{full_p_eq}).
For random position errors $\delta z \ll z$
we may re-write this as
\begin{eqnarray}
P_i &=& \int S(\nu) \cos(\phi_i + \delta \phi_i) 	~ d\nu \nonumber \\
    &=& \int S(\nu) ~ \left[ ~ \cos(\phi_i) \cos(\delta \phi_i) 
    			 - \sin(\phi_i) \sin(\delta \phi_i) ~ \right]  	~ d\nu \nonumber \\
    &=& \int S(\nu) ~ \left[ ~\cos(\phi_i) (1 - \frac{\delta \phi_i^2}{2})
    			- \delta \phi_i \sin(\phi_i)	    ~ \right]  	~ d\nu  
\label{cos_eq_2}
\end{eqnarray}
where for clarity we adopt notation
$\phi = 2 \pi \nu /c z$.
The first term is the nominal IFG amplitude
reduced by a factor
$ 1 - \delta \phi_i^2/2$.
Over many IFG realizations,
we may replace
the individual position variance $\delta z_i^2$
with the mean $< \delta z_i^2 >$.
After Fourier transformation to the frequency domain,
the resulting systematic error in the synthesized spectra
is proportional to \\
$S(\nu) (1 - <\delta \phi_i^2> /2 )$
and may be subsumed into the overall calibration.

\begin{figure}[b]
\centerline{
\includegraphics[width=3.5in]{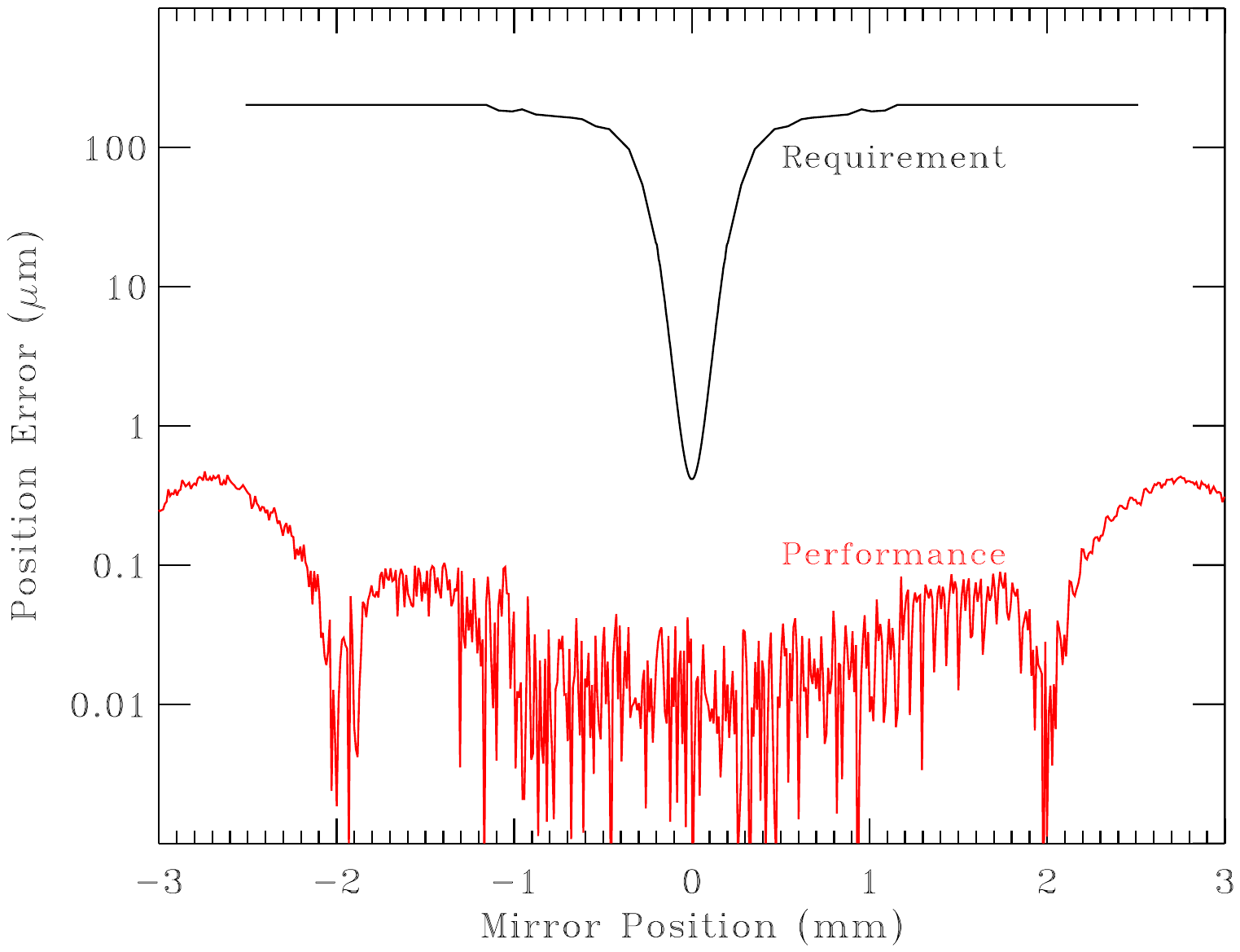}
\includegraphics[width=3.5in]{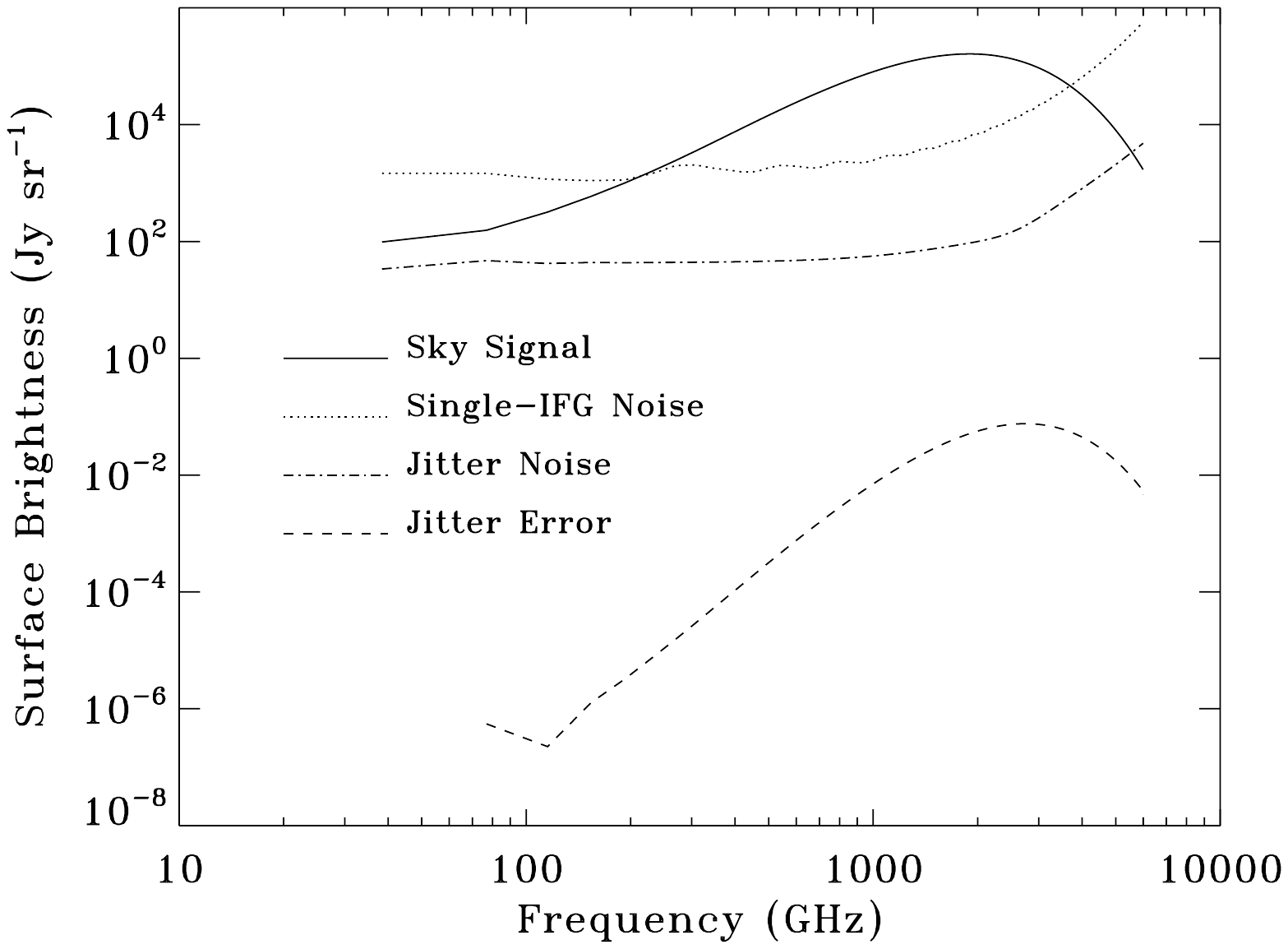}
}
\caption{
Effects of position error in the phase delay mechanism are small.
(left) Measured position errors $\delta L$ in the phase delay mechanism (
red curve)
are well below the mission requirements.
(right)
The noise term from random phase delay errors (dot-dash line)
is shown for a single IFG realization
with Gaussian position errors $\delta L = 20$~nm
($\delta z = 80$~nm).
The phase noise is small compared to the instrument white noise,
and integrates down with subsequent realizations.
The lower dashed line shows the coherent systematic error.
This term does not integrate down, 
but is subsumed into the overall calibration.
}
\label{jitter_fig}
\end{figure}

The second term varies as 
$\delta \phi_i \sin(\phi_i)$
and
sums to zero over repeated iterations of the mirror stroke.
Within any single IFG,
it represents an effective noise term.
Figure \ref{jitter_fig}
compares the phase noise
to the instrument white noise
after Fourier transformation to the frequency domain.
The left panel shows the measured position errors $\delta L$
for the PIXIE phase delay mechanism.
Including the folding of the optical path,
position error
$\delta L = 20$~nm 
corresponds to optical error
$\delta z = 80$~nm.
The right panel shows 
the phase noise term
from a single IFG realization
where
$z_i \rightarrow z_i + \delta z_i$
with
$\delta z_i$ drawn from a random Gaussian distribution
of width 80~nm.
The phase noise for a typical mid-latitude pixel
is 2 orders of magnitude below the instrument noise
for each IFG, 
and integrates down as $\sqrt{N}$ for additional samples.

\subsubsection{Spatial Interferometry}
\label{2-element-array}

When the calibrator is stowed, 
the spatial separation of the two primary mirrors
creates a phase difference $\tau = \vec{b} \cdot \vec{s} / c$
between the two parallel sky beams,
where
$\vec{s}$ is the vector from beam center to the sky
and
$\vec{b}$ is the baseline
from the center of the A-side primary to the B-side primary.
Including this effect, 
the power at each detector may be written as
\cite{nagler/etal:2015}

\begin{eqnarray}
P_{Lx} &=& \frac{1}{2} ~\int I + Q \cos(z\omega /c) + [ ~V \cos(\omega \tau) - U \sin(\omega \tau) ~] \sin(z\omega /c) ~d\omega    \nonumber \\
P_{Ly} &=& \frac{1}{2} ~\int I - Q \cos(z\omega /c) + [ ~V \cos(\omega \tau) + U \sin(\omega \tau) ~] \sin(z\omega /c) ~d\omega    \nonumber \\
P_{Rx} &=& \frac{1}{2} ~\int I + Q \cos(z\omega /c) + [ -V \cos(\omega \tau) - U \sin(\omega \tau) ~] \sin(z\omega /c) ~d\omega    \nonumber \\
P_{Ry} &=& \frac{1}{2} ~\int I - Q \cos(z\omega /c) + [ -V \cos(\omega \tau) + U \sin(\omega \tau) ~] \sin(z\omega /c) ~d\omega
\label{full_p_stokes_lag}
\end{eqnarray}
where
$V = 2 ~{\rm Im}(E_x E_y^*)$ is the Stokes parameter for circular polarization
and for clarity we have suppressed the 
notation for the optics transmission and detector efficiency.

\begin{figure}[b]
\centerline{
\includegraphics[height=4.0in]{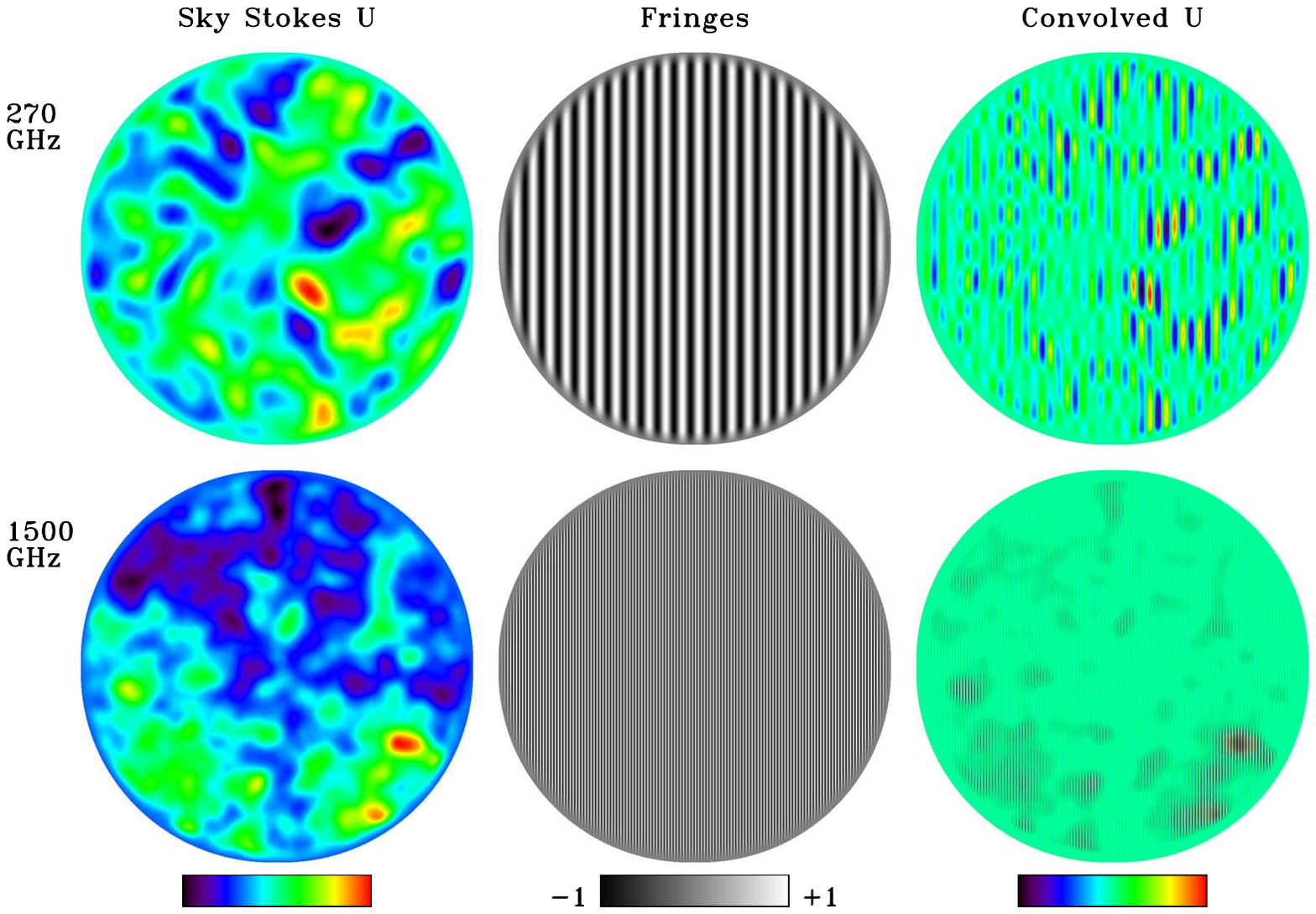}}
\caption{
The spatial separation of the two sky beams
creates a systematic error 
in the imaginary part of the Fourier transform,
proportional to Stokes $U$ component in the sky beams.
The top row shows the effect at frequency 270 GHz,
while the bottom row shows the effect at 1500 GHz.
(Left) Sky signal in Stokes $U$ for a typical mid-latitude patch
(Center) Fringe pattern from operation as a 2-element interferometer
(Right) Sky signal convolved with the fringe pattern.
The oscillatory fringe pattern across the beam
leads to nearly complete cancellation of the systematic error.
}
\label{2_beam_interferometer}
\end{figure}

Treating PIXIE as a 2-element spatial interferometer
adds two additional terms in instrument-fixed coordinates:
one term proportional to linear polarization (Stokes $U$)
and a second term proportional to circular polarization (Stokes $V$).
Both terms are modulated by
$\sin(z\omega /c)$
from the mirror movement
and thus appear only in the imaginary part of the
Fourier transform.
Several factors combine to make these ``2-element-array'' terms small.
The millimeter sky is dominated by the CMB and diffuse dust cirrus,
neither of which is thought to emit in circular polarization.
Terms proportional to Stokes $V$ can thus be neglected.
The Stokes $U$ term will be dominated by CMB E-mode polarization
and diffuse foreground emission.
This term must be integrated
over the primary beam pattern.
Since the baseline $\vec{b}$ separating the two mirrors
is normal to the antenna boresight,
the phase lag $\tau=0$
for a point source on-axis,
and oscillates as
\begin{equation}
\sin\left( \frac{2 \pi b}{\lambda} \theta \cos(\phi) \right)
\label{fringe_eq}
\end{equation}
over the beam for off-axis sources,
where 
$\theta$ is the radial distance from beam center
and
the azimuthal angle $\phi$
is defined relative to the centerline
connecting the two primary mirrors.
For mirror separation $|b| = 60$~cm,
the ratio 
$b / \lambda \gg 1$ 
at the accessible PIXIE wavelengths $\lambda > 2$~cm.

Figure \ref{2_beam_interferometer} illustrates the effect 
of 2-beam spatial interferometry.
We show the sky signal in Stokes $U$ for a 
representative mid-latitude patch
$[l,b] = [-120\deg, -50\deg]$
consisting of the CMB E-mode polarization
and the Planck model of dust polarization,
evaluated at frequencies
270 GHz near the CMB peak
and
1500 GHz near the dust peak.
The fringe pattern goes through multiple oscillations across the beam.
Integrating across the beam,
the fringe oscillations
suppress the error term
$U \sin(\omega \tau)$ 
by a factor of 40 (200,000) at 270 GHz (1500 GHz)
relative to the incident Stokes $U$ sky signal
integrated over the beam.

\section{Jackknife Tests}
\label{jackknife_section}

Jackknife tests are commonly employed by CMB missions
as a  blind test to characterize the noise
and search for residual systematic error.
Examples include
differences between the first and last half of the mission,
day vs night observations,
or data taken with and without potential noise sources present
(e.g. radio-frequency transmitters).
PIXIE can employ all of these tests;
however,
discrete symmetries within the PIXIE instrument design
and mission operations
provide additional jackknife tests
to detect, model, and remove potential systematic errors
from the data.
Table 1 lists the principal instrument symmetries
and their associated jackknife tests.

\begin{table}[t]
{
\footnotesize
\begin{center}
\begin{tabular}{| l | c | l |}
\multicolumn{3}{c}{\normalsize Table 1: Instrument Jackknife Symmetries}					\\
\hline 
Symmetry		& Time Scale	& Test	\\
\hline
Left/Right Detectors	& 4 ms		& Polarization \& optical efficiency	\\
\hline
Forward/Reverse Mirror Stroke & 4 sec	& Time constants in full readout chain	\\
\hline
Real/Imaginary FFT	& 4 sec		& $1/f$ noise, general diagnostic \\
\hline
X/Y Polarization	& 12 sec	& Detector cross-calibration		\\
\hline
Spin ($m=3-6$)		& 8-16 sec	& Beam shape, spacecraft thermal effects	\\
\hline
Spin ($m=1$)		& 48 sec	& Beam alignment	\\
\hline
A/B Optics		& 5 hrs		& Optical efficiency	\\
\hline
Hot/Cold Calibrator	& 10 hours	& Detector non-linearity	\\
\hline
Hot/Cold Optics		& 10 hours	& Stray Light	\\
\hline
Ascending/Descending Node & 6 months	& Sky Gradients, spacecraft thermal effects	\\
\hline 
Spin CW/CCW		& 2 years	& Far sidelobes \\
\hline
\end{tabular}
\end{center}
}
\label{jackknife_summary_table}
\end{table}

\subsection{Detector Pairs}
\label{detector_jackknife}

Each of the four detectors samples a unique linear combination
of signals from the $\hat{x}$ polarization in one beam
and the $\hat{y}$ polarization in the other beam
(Eq. \ref{full_p_eq}).
The  
$\hat{x}$ detector on left side of the instrument
sees the same sky signal (with a minus sign)
as the
$\hat{y}$ detector on the right side,
while the
$\hat{x}$ and $\hat{y}$ detectors sharing each concentrator
view the same sky signal
delayed 90\deg~in spacecraft spin.
The spin-delayed x-y difference
cancels sky signal to first order,
allowing detector cross-calibration.
Signals reverse sign upon the double interchange
left--right and $x$--$y$
so that the linear combinations
$(P_{Lx} + P_{Ry})/2$
and
$(P_{Ly} + P_{Rx})/2$
contain no sky signal (independent of spin).
A non-zero signal in the $L-R$ jackknife
points to
differences in the detector absorption efficiency
or
transmission through the optics.

More generally,
the four detectors have 12 possible pairwise linear combinations
(Appendix A).
Jackknife comparison of detector pairs
tests both the polarized detector absorption efficiency $\epsilon$ 
as well as the transmission $f$ through the A- and B-side optics
from the detector to the sky (or calibrator).
Since both $\epsilon$ and $f$ are defined relative
to the external blackbody calibrator,
any differences
$\epsilon_i - \epsilon_j$ 
or
$f_A - f_B$
appear at second order in the sky data
and may be identified and corrected at this level
by the pairwise detector comparison
($\S$\ref{a_vs_b_subsection}).

Differences in the beam patterns
can also produce detector-specific error signals
\cite{pixie_4_port_2019}.
These can be distinguished from
differences in detector absorption efficiency or optical transmission
by their dependence on spacecraft spin ($\S$\ref{spin_subsection}).

\begin{figure}[t]
\centerline{
\includegraphics[width=6.5in]{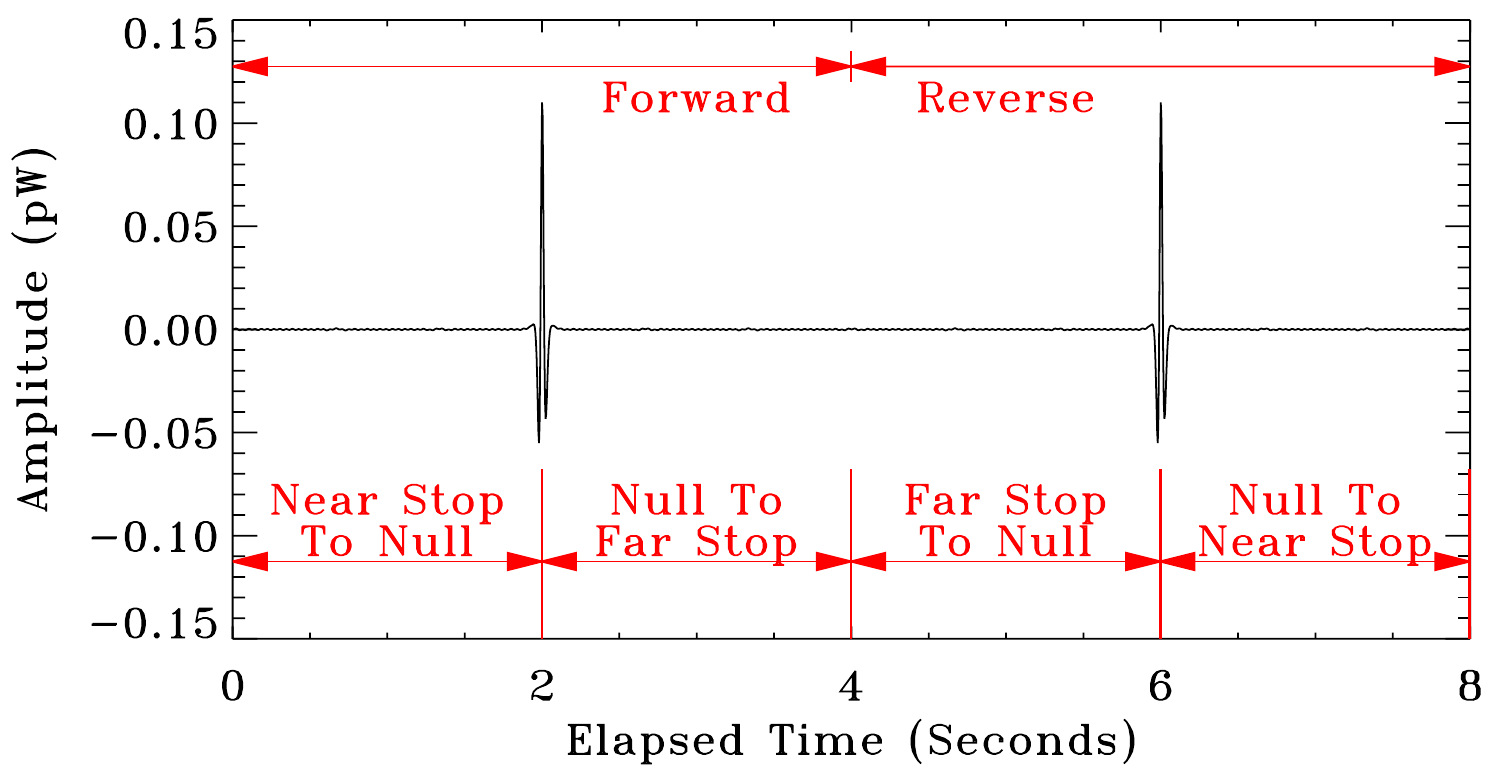}}
\caption{
The phase delay mirror stroke provides 4 copies of the sky information
with different spatial/temporal symmetries.  
Jackknife tests exploiting these symmetries correct for
systematic errors resulting from 
the mirror zero position offset
and detector time constants. 
}
\label{mirror_stroke}
\end{figure}

\subsection{Mirror Stroke}
\label{mtm_section}

The ideal interferograms on each detector depend only on the
absolute value of the optical phase delay,
or equivalently,
the absolute mirror displacement from zero path difference.
The full mirror stroke,
from the far stop through zero path difference 
to the near stop and then back again,
provides 4 copies of the sky data
with different space/time symmetries
(Figure \ref{mirror_stroke}).

\begin{figure}[b]
\centerline{
\includegraphics[width=3.5in]{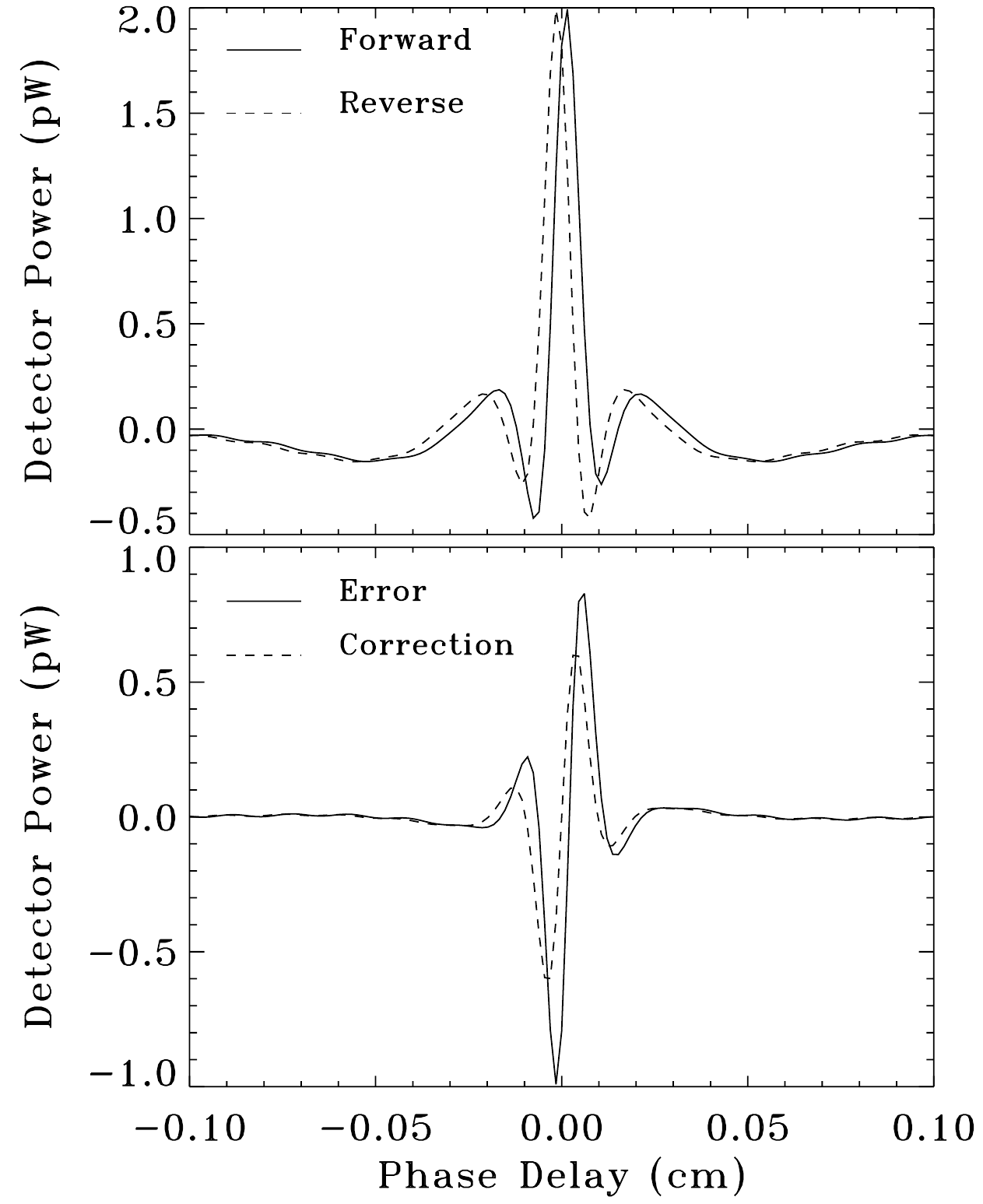}}
\caption{
(Top) Interferograms from the forward and reverse mirror directions
show the delay from the detector time constant.
(Bottom) The anti-symmetric component of the forward/reverse mirror strokes
provides a first-order correction to the error 
between the measured IFGs and the true sky signal.
}
\label{mirror_jackknife}
\end{figure}

Time reversal symmetry provides a straightforward evaluation
of time constants in the readout chain.
This includes both the thermal time constant(s)
intrinsic to each detector,
as well as time constants or filters
employed in the post-detector readout electronics.
Figure \ref{mirror_jackknife} shows an example.
Detector time constants act as a single-pole filter,
with error signal proportional to the time derivative of the incident signal
\begin{equation}
P_i \rightarrow P_i + \alpha P_{i-1} ~,
\label{time_comst_eq}
\end{equation}
where
$P_i$ refers to the 
power at the detector for the $i^{\rm th}$ time-ordered sample.
We first generate a set of interferograms for a full mirror stroke,
using as inputs a sky model
with CMB, synchrotron, dust continuum spectra
plus the bright C{\sc{ii}} and CO lines,
with the calibrator held 0.3~mK above the CMB.
We then process the time-ordered IFG from the true sky signals
through a single-pole filter with 256~Hz sampling
and detector time constant 7~ms
to create as-observed IFGs for the forward and reverse mirror strokes.
The top panel of Figure \ref{mirror_jackknife}
compares the resulting forward and reverse IFGs.

The delay induced by the detector time constant
causes the observed IFGs to differ from the true sky signal
(bottom panel of Figure \ref{mirror_jackknife}).
Sorting the IFGs by mirror direction
identifies the resulting systematic error
and provides a first-order correction.
The (forward + reverse) IFG sum is symmetric about zero path difference,
and provides an initial estimate for the sky signal.
The (forward-reverse) IFG difference is anti-symmetric about ZPD
and
provides a first-order correction.
A two-year mission provides 15~million repetitions of the mirror stroke
for each detector.
Since the Fourier-transformed sky spectra
are referred to the calibrator Planck spectrum,
residual errors after correction
appear at second order in the calibration
and at third order in the sky spectra.

The mirror stroke also
enables a determination
of the mirror position at zero path difference.
Physical differences in the 
optical path  between the two sides of the instrument
as well as 
offsets in the mirror position readout
can create a systematic difference $\Delta z$
between the mirror location at commanded position $z=0$
and the actual mirror location at the white-light peak.
Let
\begin{eqnarray}
P_s(z) &=& \frac{1}{2} \left[ P_F(z) + P_R(z) \right]	\\
P_d(z) &=& \frac{1}{2} \left[ P_F(z) - P_R(z) \right]	
\label{stroke_sum_diff}
\end{eqnarray}
represent the symmetric sum
and anti-symmetric difference
of the forward and reverse mirror strokes,
sorted by mirror position.
Since true sky signals obey
$P_s(z) = P_s(-z)$ 
for all mirror positions $z$,
the jackknife
$P_s(z) - P_s(-z)$
tests for fixed phase errors in $z$.

\subsection{Real and Imaginary Fourier Transform}
\label{real_vs_imaginary_fft}

The Fourier transform of the sampled interferograms
produces frequency spectra $I(\nu)$
with both real and imaginary terms.
By symmetry,
only the real part contains sky signals,
while the imaginary part 
contains
an independent realization of the noise 
plus systematic errors
from any effects that are not symmetric 
with respect to mirror position $|z|$.
This includes mirror position errors from 
ZPD offsets
($\S$\ref{MTM_jitter_section})
as well as
transient position errors excited as the mirror
reverses direction at the end of each sweep.
Systematic errors in mirror position 
resulting from the mirror turnaround
are identified by comparing
the IFG at start of each sweep
to data at the end of the preceding sweep
(when transport transients have decayed away).
Although the resulting signal may be identified and modeled
using the imaginary transform,
in practice
it is simpler to discard samples near mirror turnaround
as part of the IFG apodization.
Since data far from ZPD primarily encode line emission,
the effect is a few-percent decrease in line sensitivity
with no significant effect on continuum spectra.

All data products produced from the real Fourier transform
(sky maps, frequency spectra, power spectra, etc)
can also be generated from the imaginary transform.
Since both the real and imaginary Fourier outputs
are generated from the {\it same} input data set,
the resulting noise estimates in the imaginary part
are present at full mission sensitivity
and include all effects of
sampling, calibration, and data processing.

\subsection{Spin}
\label{spin_subsection}

Sorting the observations within each pixel by spacecraft spin angle
allows identification and subtraction of additional systematic errors.

\subsubsection{Spin Moments}
\label{spin_jackknife}

The spin-independent $m=0$ mode
contains the unpolarized sky spectra (Stokes I).
Regardless of the individual beam shapes,
the $m=0$ response
is azimuthally symmetric
(although it may contain radial information)
\cite{pixie_4_port_2019}.

Angular offsets between the $A$ and $B$ beams on the sky
couple to
spatial gradients on the sky
to create a dipolar $m=1$ response
($\S$\ref{beam_section}).
Data taken with the calibrator deployed to block the A beam
determine the offset of the B (sky) beam,
and conversely for data with the calibrator blocking the B beam.
With the calibrator stowed,
both beams are open to the sky
to allow independent determination of the beam offsets.

The $m=2$ term contains
the sky polarization signal
as well as temperature-polarization systematic error coupling
from beam ellipticity.
Common-mode ellipticity cancels in the A-B optical differencing,
leaving only the differential A-B ellipticity.
The mirror symmetry (Figure \ref{symmetry_cartoon})
minimizes the differential ellipticity.
After accounting for machining and assembly tolerances,
the differential ellipticity is below $3 \times 10^{-3}$
for individual detectors.
Linear combinations of detectors
can be analyzed with respect to spin angle
to further probe the differential beam patterns
\cite{pixie_4_port_2019}.

Higher order spin terms do not couple to either intensity or polarization.
The measured response can be compared to predictions based on beam models
as a blind test for the beam shape.
Thermal or electrical effects
(either additive or multiplicative)
sourced by the three-fold symmetry of the solar panels
occur at $m=3$.

\subsubsection{Clockwise vs Counterclockwise}

Azimuthal structure in the far sidelobes 
couples to gradients in the sky
(primarily of Galactic origin)
to source signal variation at $m=1$.
The spacecraft spin about the beam axis
reverses each year,
with one year in clockwise rotation
followed by a second year in counterclockwise rotation.
Reversing the spin direction
reverses the time-ordered $m=1$ signal
while preserving the desired $m=0$ and $m=2$ signals.
Signals sourced by solar heating of the spacecraft
are likely to have a thermal phase lag
with respect to the solar angle.
Reversing the spacecraft spin
reverses the phase lag.

\subsection{A vs B Optics}
\label{a_vs_b_subsection}

Signals entering through the A and B optics 
appear with opposite signs on each detector
(Eq. \ref{full_p_eq}).
Data taken with the calibrator 
covering the A beam
vs the B beam
provide in-flight measurement of the 
transmission through the optics.
We sort the data from each sky pixel
and
sum the data taken with the calibrator blocking the A beam
with
those from the same sky pixel
but with the calibrator blocking the B beam.
We thus have 
\begin{eqnarray}
\Delta_{\rm cal} &=& P_{Lx}({\rm Cal,A}) + P_{Lx}({\rm Cal,B}) 	\nonumber	\\
		 & \propto & 
	\epsilon_{Lx} \left [f_A E_{x,{\rm cal}}^2 - f_B E_{y,{\rm sky}}^2 \right]
   +
	\epsilon_{Lx} \left [f_A E_{x,{\rm sky}}^2 - f_B E_{y,{\rm cal}}^2 \right]
\label{AB_sum_1}
\end{eqnarray}
for the single left $\hat{x}$ detector,
with similar expressions for the other detectors.
Performing the Fourier transform to derive the frequency spectra, 
we may write
\begin{eqnarray}
\Delta_{\rm cal}(T_A, T_B) = \epsilon_{Lx}  ~[
	&~&  M( B_\nu(T_A) - B_\nu(T_B) )
	 + D( B_\nu(T_A) + B_\nu(T_B) )		\nonumber \\
	&+& M \, \boldsymbol{Q}_{\rm sky}(\nu)
	 - D \, \boldsymbol{I}_{\rm sky}(\nu)	~]  
\label{AB_sum_2}
\end{eqnarray}
where
$T_A$ and $T_B$ are the calibrator temperature 
when deployed over the A and B beams\footnote{
By definition, a blackbody is unpolarized so that 
$E_{x,{\rm cal}}^2 = E_{y,{\rm cal}}^2$.
},
$M=(f_A + f_B)/2$ is the mean transmission,
$D = (f_A - f_B)/2$ is the difference in transmission
between the A- and B-side optics,
$\boldsymbol{I}(\nu) = E_x^2 + E_y^2$ is the sky intensity (Stokes I)
and
$\boldsymbol{Q}(\nu) = E_x^2 - E_y^2$ is the sky polarization (Stokes Q).
We then evaluate the double difference
comparing data 
when the calibrator over side A 
is held at a fixed temperature offset $\Delta T$ 
warmer than when over side B,
to the equivalent data (on the same sky pixel)
when the calibrator over side B 
is held at the same fixed $\Delta T$ 
warmer than side A,
\begin{eqnarray}
\Delta_{\rm cal}^{(1)}(\Delta T) &=& 
	\Delta_{\rm cal}(T+\Delta T, T) - \Delta_{\rm cal}(T, T + \Delta T)  \nonumber \\
  &=&	2  M \frac{\partial B}{\partial T} \Delta T ~.
\label{AB_mean_eq}
\end{eqnarray}
The sky signals cancel,
as does the term proportional 
to the mean calibrator power times the differential transmission,
allowing a clean determination of the mean transmission through the optics.
This provides the fundamental calibration
to convert the digitized telemetry signals
to physical units (e.g. Jy sr$^{-1}$)\cite{kogut/fixsen:2019}.
Similarly,
we may hold the calibrator at the same temperature $T_A=T_B$
over both beams,
and evaluate the double difference
when the common temperature is increased by $\Delta T$,
%
%
\begin{eqnarray}
\Delta_{\rm cal}^{(2)}(\Delta T) &=& 
	\Delta_{\rm cal}(T, T) - \Delta_{\rm cal}(T + \Delta T, T + \Delta T)  \nonumber \\
  &=&	2  D \left( ~B_\nu(T) +  \frac{\partial B}{\partial T} \Delta T ~\right) ~.
\label{AB_diff_eq}
\end{eqnarray}
This provides a clean determination of the differential transmission
through the optics,
independently for each of the 4 detectors.
When combined with the pairwise detector jackknife tests
(Appendix A),
both the optics transmission $f$ 
and the detector absorption efficiency $\epsilon$
can be measured and tracked throughout the flight.

\subsection{Hot vs Cold Calibrator}

Data with the calibrator deployed provide 
in-flight measurements of the detector non-linearity.
The loading on an individual detector is dominated by the CMB monopole.
Changes in the loading from the sky signal variation
are dominated by the 3~mK CMB dipole,
which changes the mean loading by 0.25\%.
At the start of each great-circle scan,
the calibrator is
moved to a new position
and
set to a new temperature.
The temperature setpoints span the range [2.720, 2.730]~K
to bracket the CMB monopole at 2.725~K.
Calibration thus varies the detector loading by $\pm 0.7$\%
so that the detector performance
is always interpolated between values measured in flight.
Calibrator temperature excursions of $\pm$5 mK
from a single great-circle comparison
provide a signal-to-noise ratio 
above $10^3$
within individual frequency channels,
allowing any changes in detector performance to be monitored 
throughout flight\cite{kogut/fixsen:2019}.

\subsection{Hot vs Cold Optics}

Losses in the optics terminate within the instrument
and are replaced by photons emitted at the instrument temperature.
Each of the surfaces within the instrument
is individually controlled in temperature.
The set of temperature setpoints is updated twice per great-circle scan
to provide a unique temperature profile vs time
for each surface
($\S$\ref{internal_emission}),
which can be compared to the detector total power
to determine the effective optical coupling from the detector to each surface.
Stray-light signals that pass through the mirror phase delay
change sign when the surface is warmer vs colder 
than the sky/calibrator temperature,
allowing a second measure of their coupling
through corresponding changes in the interferograms.

\subsection{Ascending vs Descending Node}

The spacecraft scans the beams through a great circle
in ecliptic coordinates,
perpendicular to the sun line.
As the beams move across the sky,
hot and cold spots on scales much smaller than the beams
enter and exit the field of view,
creating a source of signal variation
on time scales comparable to or shorter than the mirror stroke.
Signal variation on short time scales
correspond to high spatial frequencies in the interferograms
and 
are Fourier transformed into 
systematic error terms frequencies above 6 THz in the sky spectra 
\cite{naess/etal:2019}.
Since the signal variations as small-scale anisotropies
enter and exit the beams
are not symmetric with respect to the mirror stroke,
the resulting systematic error signals
appear in both the real and imaginary parts of the Fourier transform.
Noise models derived from the imaginary part of the Fourier transform
($\S$\ref{real_vs_imaginary_fft})
include these terms,
which are small compared to either the instrument noise
or sky signals
\cite{naess/etal:2019}.

Every six months,
the beams retrace the same great circle on the sky
but in the opposite direction,
reversing the time sequence 
from small-scale anisotropy entering and exiting the beams.
Differencing the interferograms
from the ascending vs descending node observations
within each pixel
provide a second way to isolate and quantity 
the noise contribution from small-scale anisotropy.

\section{Discussion}

Minimization, identification, and correction
for potential systematic errors
play a critical role for precision measurements
of CMB spectral distortions or polarization.
In many respects PIXIE represents a modernized version
of the seminal FIRAS instrument,
whose measurements of the CMB spectrum
are still unsurpassed some 30 years later.
PIXIE improves the FIRAS design in several significant ways.
PIXIE's colder detector temperature (0.1~K vs 1.4~K)
allows operation near the limit imposed by photon statistics,
while the larger etendue
(4~cm$^2$~sr vs 1.5~cm$^2$~sr per detector)
allows more photons to be detected.
The FIRAS measurements were not limited by systematic effects,
but by the instrument noise during the limited integration time 
spent observing the external calibrator
through the single sky horn
\cite{firas_syserr_1994}.
PIXIE's 2-beam design allows the sky and external calibrator
to be observed simultaneously,
ensuring equal integration time
for both sky and calibrator.
Additional design modifications 
provide the necessary control of systematic effects
to take advantage of the improved sensitivity.
FIRAS maintained the sky horn and calibrator at 2.725~K
but held the FTS and detectors at temperature 1.4~K,
1.3~K below the CMB or external calibrator.
PIXIE closely approximates a blackbody cavity,
with all optical components except the detectors
maintained within 5~mK of the CMB temperature.
FIRAS interfered a single sky beam against an internal calibrator,
while PIXIE's fully symmetric design
interferes two sky beams
including all instrument optical elements.
The differential optics provide pre-detection cancellation 
of potential systematic errors
while
enabling a number of jackknife tests
to identify, model, and correct residual errors.

\begin{figure}[b]
\centerline{
\includegraphics[width=5.0in]{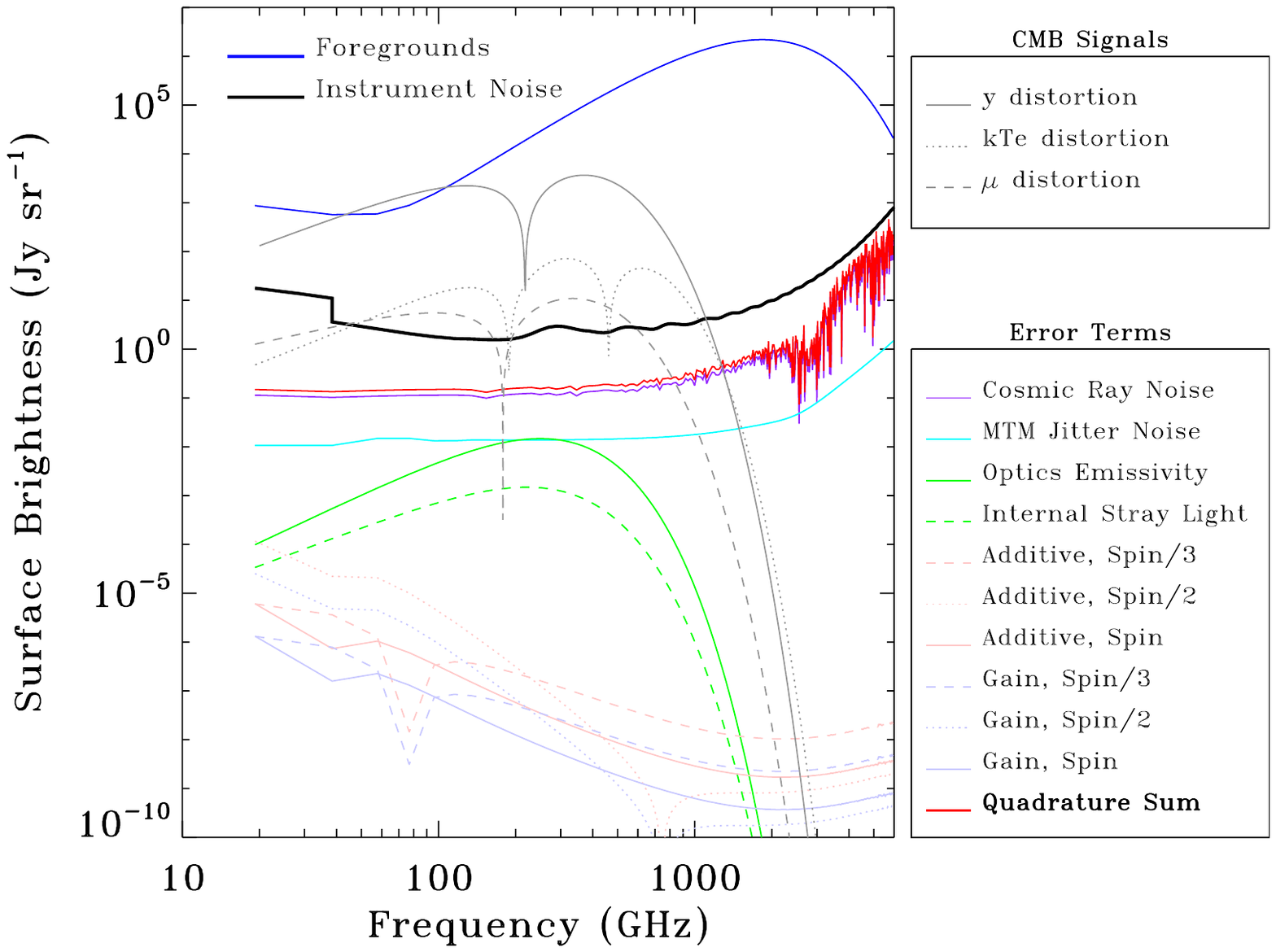}}
\caption{
PIXIE systematic error budget for spectral distortions.
Individual curves show the residual systematic error terms,
after correction,
and the quadrature sum.
The combined systematic errors
are small compared to cosmological signals
or the baseline 2-year mission noise.
}
\label{syserr_sum_spectra}
\end{figure}

Figure \ref{syserr_sum_spectra} summarizes the systematic error budget
for measurements of CMB spectral distortions.
The two largest effects,
cosmic ray hits
and position jitter in the phase-delay mechanism,
create additional noise terms
at levels 10 to 100 times below the instrument noise,
but do not inject coherent spectral signals.
Since these are noise terms,
they integrate down with observing time
and are shown in Fig \ref{syserr_sum_spectra}
for the nominal two-year mission.
The largest spectral signature,
from assumed 1\% differential emissivity within the beam-forming optics
($\S$\ref{internal_emission}),
is two orders of magnitude below the instrument noise
and three orders of magnitude below
the cosmological $\mu$ distortion.
All other effects are smaller still.
We normalize the amplitude for additive (post-detection) signals
by scaling the results of $\S$\ref{periodic_section},
using sinusoidal signals
with amplitude equal to 10\% of the instrument noise
integrated over a single spacecraft spin period
(so as to avoid detection in a single spin).
We normalize the amplitude of spin-modulated gain variation at
$\delta G/G = 10^{-4}$,
set by the upper limit of 
10~mK thermal variation of the instrument electronics
at the spacecraft spin period
($\S$\ref{gain_error}).
Both effects contribute negligibly to the systematic error budget.

\begin{figure}[b]
\centerline{
\includegraphics[width=5.0in]{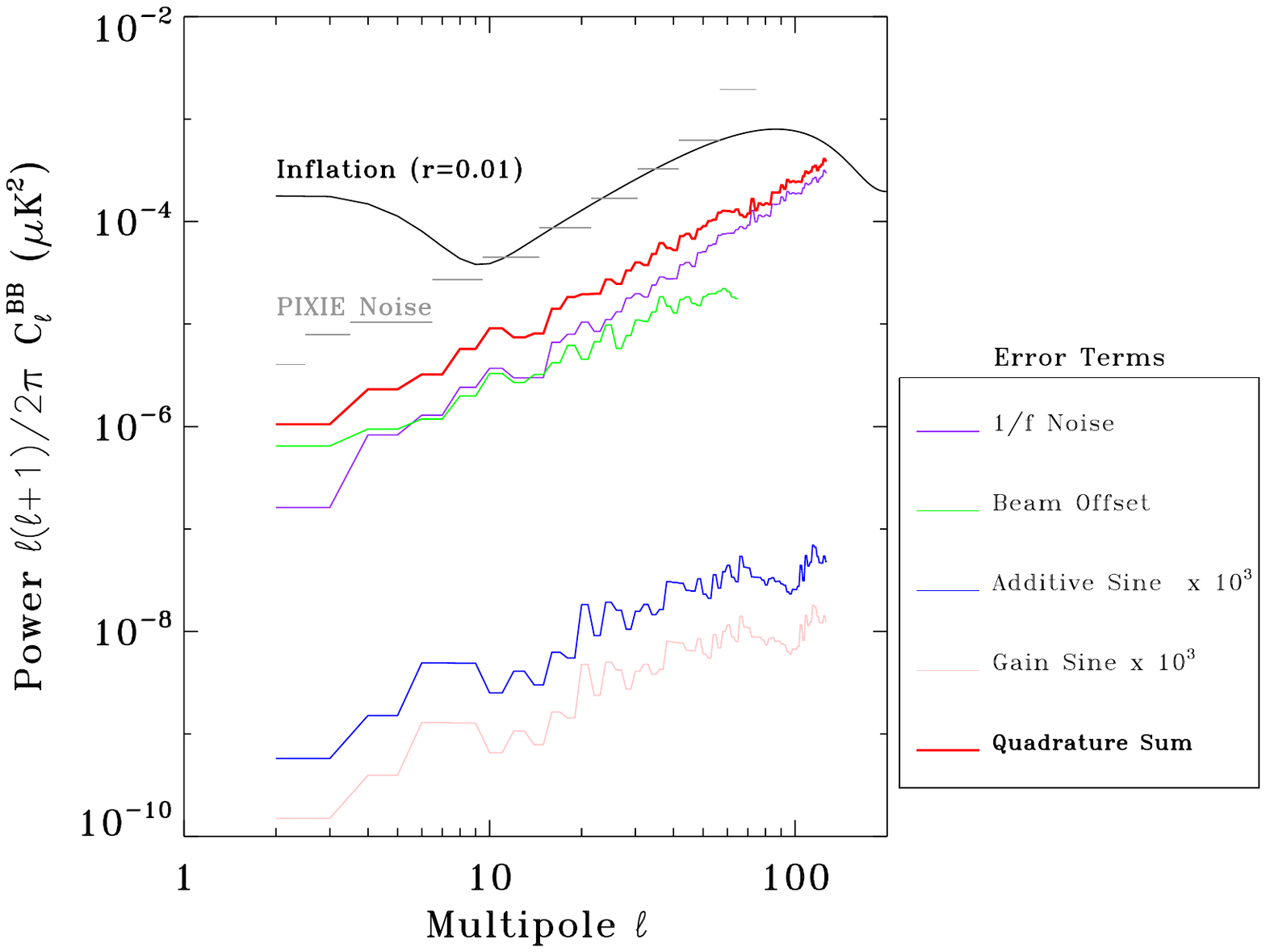}}
\caption{
PIXIE systematic error budget for B-mode polarization.
Individual curves show the residual systematic error terms,
after correction,
and the quadrature sum.
The combined systematic errors
are small compared to cosmological signals
or the baseline 2-year mission noise.
}
\label{syserr_sum_cl}
\end{figure}

Figure \ref{syserr_sum_cl} summarizes the systematic error budget 
for B-mode polarization.
We use both the spectral and spatial distribution
of each systematic error signal
to quantify the expected contribution of each source
to B-mode polarization in the CMB.
For each error term,
we first simulate time-ordered data 
to generate the Stokes IQU parameters
in each spatial pixel and each frequency channel
for the full two-year mission.
We then fit the frequency spectra
within each pixel
to the CMB $\partial B / \partial T$ anisotropy spectrum
to derive the amplitude 
of the component mimicking CMB anisotropy in each Stokes component.
Finally,
we use the HEALPIX {\tt anafast} code
to derive the B-mode power spectrum
from the resulting Stokes Q and U maps in the fitted CMB component
at Galactic latitudes $|b| > 20\deg$.
The largest term
($1/f$ noise with knee frequency 0.1~Hz)
is an order of magnitude below the white noise contribution,
but does not contribute coherent structure to the power spectrum.
The next largest term
is the residual error from beam offsets
after fitting the 2-year mission
for the individual offsets in the A and B beams
for each detector
($\S$\ref{beam_section}).
An additive sine wave 
with amplitude 10\% of the white-noise level
at twice the spin frequency
lies over 6 orders of magnitude below the noise.
Systematic errors from 
gain variations 
$\delta G/G = 10^{-4}$
at twice the spin frequency
are smaller still.
All systematic error terms
have white-noise angular power spectra
($C_\ell$ independent of $\ell$)
which further distinguishes them
from the inflationary B-mode signal
on large angular scales.

It is important to note that PIXIE's systematic error suppression
does not rely on any single cancellation
to achieve the mission goals,
but instead chains together multiple symmetries
to successively reduce potential errors.
PIXIE closely approximates a blackbody cavity,
maintaining the calibrator,
optics,
and absorbing walls
within a few mK of the CMB sky temperature.
The largest possible instrumental spectral signature 
is thus of order a few mK,
which is reduced to tens of $\mu$K
by typical percent-level mirror emissivity.
Any spectral signatures generated on one side of the instrument
are then differenced against comparable signals 
from the other side. 
Percent-level differences in the optical symmetry
then leave residuals at the few-hundred nK level.
Active control of the optics temperatures
periodically reverses the sign of any residual spectral signature.
The weighting of hot-vs-cold observations
can be adjusted in post-flight analysis
to match at percent levels or better,
reducing the instrumental signal to nK levels or below.
A similar chain suppresses instrumental signals in polarization.
The Fourier transformation maps additive signals
with periods longer than the mirror stroke
onto the lowest synthesized frequency channels,
effectively isolating such signals
from the amplitude-modulated response
to true sky signals.
The multi-moded optics
illuminate a single spatial pixel on the beam axis,
producing a tophat beam with ellipticity 7\%.
Sky signals from the two beams are differenced optically
so that only the differential ellipticity couples
to large-scale gradients on the sky.
The left-right instrument symmetry 
matches the $\hat{x}-\hat{y}$ polarization difference
to few-percent accuracy,
reducing temperature-polarization coupling by a comparable amount.
Each concentrator contains two independent detectors
sensitive to orthogonal polarization states,
which view the same sky pixel through the same optics.
The differential ellipticity thus cancels
to second order in the sum or difference comparison
of different detectors.
The spacecraft provides uniform sampling 
of  the sky signal vs spin angle for each pixel
over the full $2\pi$ range,
cleanly separating the $m=2$ sky polarization
from the dominant $m=1$ beam effects.
The mirror stroke and spacecraft spin
are fast compared to the great-circle scan motion 
of the beams across the sky,
so that the polarization state 
is determined independently in each pixel,
eliminating pixel-pixel covariance 
on angular scales larger than the beam.

PIXIE's multiple levels of signal modulation
maintain approximate orthogonality
between sky signals
and systematic errors.
Signal degeneracy requires that
a given systematic error signal
({\it e.g.}, internal emission within the instrument)
have the same variation in the time domain
as either the sky signal
or a competing systematic error term.
The Fourier transform spectrometer
imposes a complex amplitude modulation
on true sky signals,
which are further modified by the
time-dependent motion of spacecraft spin,
great-circle scan,
and annual orbital precession.
None of the systematic errors considered here
include all of these modulations,
while additional modulations imposed on the instrument
but not the sky signals
({\it e.g.}, periodic steps in the instrument temperatures)
serve to further separate and identify systematic errors.
During its two-year baseline mission,
PIXIE will accumulate $1.5 \times 10^{10}$
samples of the interferograms.
The calibration and mapping pipeline
will compress this raw time-ordered data
into $1.5 \times 10^{8}$ data points
in the combined real and imaginary sky cubes.
Fitting additional terms to identify and subtract
the systematic error terms
produces a negligible decrease in the total degrees of freedom,
and a correspondingly small effect on the projected sky noise.

\section{Conclusions}

We use detailed time-ordered simulations
to evaluate the amplitude of systematic error signals
in the PIXIE data.
PIXIE combines multiple levels of 
null operation,
signal modulation,
and differencing
to reduce systematic errors to negligible levels.
Jackknife tests
based on discrete instrument symmetries
provide an independent means to
identify, model, and remove remaining instrumental signals.
The Fourier transform spectrometer
modulates the sky signal
over time scales 4~ms to 4~s.
Signal drifts or post-detection pickup
on longer time scales are mapped 
to the lowest synthesized frequency channels
and do not project 
to the frequency or spatial distribution
of either the CMB or astrophysical foregrounds.
The largest systematic error signals,
after identification and correction,
are the residual noise terms
from cosmic-ray hits to the detector absorber
and the $1/f$ response of the electronics.
Both appear in the spectral and polarization data
as effective white-noise terms
at levels of a few percent of the dominant photon noise contribution.
Coherent instrumental effects,
which do not integrate down with observing time,
are smaller still.
The largest effect for spectral distortions
is the differential emissivity of the beam-forming optics.
Percent-level matching of the optical properties
leaves a residual systematic error
with amplitude 0.1\% of the 
cosmological $\mu$-distortion
from Silk damping of primordial density perturbations.
Other effects in the frequency spectra are smaller still.
Systematic errors for CMB polarization
are suppressed in both the spectral and spatial dimensions.
As with spectral distortions,
the largest effect is the contribution from instrumental $1/f$ noise,
which appears in the polarization maps 
as a white-noise term
with amplitude one percent of the photon noise.
The largest coherent effect for polarization
is temperature-polarization coupling
induced by offsets between the spacecraft spin axis
and the individual beam boresights.
The spacecraft spin allows identification
and correction for beam offsets;
the residual after correction
has amplitude below 1\% of the integrated instrument noise.
Systematic errors in polarization
have a white noise power spectrum
($C_\ell$ independent of $\ell$)
and are readily distinguished from cosmological signals
on large angular scales.

\clearpage

\appendix
\section{Pairwise Detector Comparison}

The 4 detectors provide 12 distinct pairwise linear combinations.
Expanding Eq. \ref{full_p_eq}, we have
\begin{eqnarray}
P_{Lx} + P_{Ly} &=&   ~~(\epsilon_{Lx}-\epsilon_{Ly})M\boldsymbol{Q} ~+~ (\epsilon_{Lx}+\epsilon_{Ly})D\boldsymbol{I} \\
P_{Lx} - P_{Ly} &=&   ~~(\epsilon_{Lx}+\epsilon_{Ly})M\boldsymbol{Q} ~+~ (\epsilon_{Lx}-\epsilon_{Ly})D\boldsymbol{I} \\
P_{Lx} + P_{Ry} &=&   ~~(\epsilon_{Lx}-\epsilon_{Ry})M\boldsymbol{Q} ~+~ (\epsilon_{Lx}-\epsilon_{Ry})D\boldsymbol{I} \\
P_{Lx} - P_{Ry} &=&   ~~(\epsilon_{Lx}+\epsilon_{Ry})M\boldsymbol{Q} ~+~ (\epsilon_{Lx}+\epsilon_{Ry})D\boldsymbol{I} \\
P_{Lx} + P_{Rx} &=&   ~~(\epsilon_{Lx}+\epsilon_{Rx})M\boldsymbol{Q} ~+~ (\epsilon_{Lx}-\epsilon_{Rx})D\boldsymbol{I} \\
P_{Lx} - P_{Rx} &=&   ~~(\epsilon_{Lx}-\epsilon_{Rx})M\boldsymbol{Q} ~+~ (\epsilon_{Lx}+\epsilon_{Rx})D\boldsymbol{I} \\
P_{Ly} + P_{Rx} &=&   ~~(\epsilon_{Rx}-\epsilon_{Ly})M\boldsymbol{Q} ~-~ (\epsilon_{Rx}-\epsilon_{Ly})D\boldsymbol{I} \\
P_{Ly} - P_{Rx} &=&  -(\epsilon_{Rx}+\epsilon_{Ly})M\boldsymbol{Q} ~+~ (\epsilon_{Rx}+\epsilon_{Ly})D\boldsymbol{I} \\
P_{Ly} + P_{Ry} &=&  -(\epsilon_{Ly}+\epsilon_{Ry})M\boldsymbol{Q} ~+~ (\epsilon_{Ly}-\epsilon_{Ry})D\boldsymbol{I} \\
P_{Ly} - P_{Ry} &=&   ~~(\epsilon_{Ry}-\epsilon_{Ly})M\boldsymbol{Q} ~+~ (\epsilon_{Ly}+\epsilon_{Ry})D\boldsymbol{I} \\
P_{Rx} + P_{Ry} &=&   ~~(\epsilon_{Rx}-\epsilon_{Ry})M\boldsymbol{Q} ~-~ (\epsilon_{Rx}+\epsilon_{Ry})D\boldsymbol{I} \\
P_{Rx} - P_{Ry} &=&   ~~(\epsilon_{Rx}+\epsilon_{Ry})M\boldsymbol{Q} ~-~ (\epsilon_{Rx}-\epsilon_{Ry})D\boldsymbol{I}
\label{det_pair_array}
\end{eqnarray}
where
$M = (f_A + f_B)/2$ is the mean transmission,
$D = (f_A - f_B)/2$ is the difference in transmission
between the A- and B-side optics,
$\boldsymbol{I} = E_x^2 + E_y^2$ is the intensity (Stokes I)
and
$\boldsymbol{Q} = E_x^2 - E_y^2$ is the polarization (Stokes Q).
Both the polarized detector absorption efficiency $\epsilon$
and the transmission efficiency $f$ 
from the detector to the sky
are expected to be slowly-varying functions of frequency.

\clearpage

%
%
%

\providecommand{\href}[2]{#2}\begingroup\raggedright\endgroup

\end{spacing}
\end{document}